\documentclass[10pt,a4paper,journal]{IEEEtran}
\usepackage{cite,graphicx,psfrag,amsmath,color,epsfig}
\newcommand{\gt}{\tau}

\def\bm#1{\mbox{\boldmath $#1$}}
\newcommand{\vgt}{\mbox{$\bm \gt$}}
\newcommand{\vphi}{\mbox{$\bm \phi$}}
\newcommand{\bsigma}{{\bm\sigma}}
\newcommand{\bSigma}{{\boldsymbol{\Sigma}}}
\newcommand{\bLambda}{{\boldsymbol{\Lambda}}}
\newcommand{\btheta}{{\boldsymbol{\theta}}}

\newcommand{\diag}{{\ensuremath{\mathrm{diag}}}}
\newcommand{\vect}{{\ensuremath{\mathrm{vec}}}}
\newcommand{\jcmplx}{{\ensuremath{\mathrm{j}}}}
\makeatletter
\def\argmin{\mathop{\operator@font arg\,min}}
\makeatother

\newcommand{\va}{{\ensuremath{{\mathbf{a}}}}}
\newcommand{\vg}{{\ensuremath{{\mathbf{g}}}}}
\newcommand{\vk}{{\ensuremath{{\mathbf{k}}}}}
\newcommand{\vp}{{\ensuremath{{\mathbf{p}}}}}
\newcommand{\vw}{{\ensuremath{{\mathbf{w}}}}}
\newcommand{\vx}{{\ensuremath{{\mathbf{x}}}}}
\newcommand{\vy}{{\ensuremath{{\mathbf{y}}}}}

\newcommand{\ba}{{\ensuremath{{\mathbf{a}}}}}
\newcommand{\bac}{{\ensuremath{{\mathbf{\bar{a}}}}}}
\newcommand{\bg}{{\ensuremath{{\mathbf{g}}}}}
\newcommand{\bn}{{\ensuremath{{\mathbf{n}}}}}
\newcommand{\bp}{{\ensuremath{{\mathbf{p}}}}}
\newcommand{\bw}{{\ensuremath{{\mathbf{w}}}}}
\newcommand{\bx}{{\ensuremath{{\mathbf{x}}}}}
\newcommand{\bz}{{\ensuremath{{\mathbf{z}}}}}

\newcommand{\mC}{{\ensuremath{\mathbf{C}}}}
\newcommand{\mG}{{\ensuremath{\mathbf{G}}}}
\newcommand{\bRh}{{\ensuremath{\mathbf{\hat{R}}}}}
\newcommand{\mU}{{\ensuremath{\mathbf{U}}}}
\newcommand{\mW}{{\ensuremath{\mathbf{W}}}}
\newcommand{\mZ}{{\ensuremath{\mathbf{Z}}}}

\newcommand{\bA}{{\ensuremath{\mathbf{A}}}}
\newcommand{\bC}{{\ensuremath{\mathbf{C}}}}
\newcommand{\bD}{{\ensuremath{\mathbf{D}}}}
\newcommand{\bG}{{\ensuremath{\mathbf{G}}}}
\newcommand{\bI}{{\ensuremath{\mathbf{I}}}}
\newcommand{\bK}{{\ensuremath{\mathbf{K}}}}
\newcommand{\bR}{{\ensuremath{\mathbf{R}}}}
\newcommand{\bU}{{\ensuremath{\mathbf{U}}}}

\title{Calibration Challenges for the Next Generation of Radio Telescopes}
\author{Stefan J. Wijnholds, Sebastiaan van der Tol, Ronald Nijboer and
  Alle-Jan van der Veen}

\def\revdate{accepted by SPM, final version, submitted \today}
\begin{document}

\maketitle
\IEEEpeerreviewmaketitle

Instruments for radio astronomical observations have come a long way. While
the first telescopes were based on very large dishes and 2-antenna
interferometers, current instruments consist of dozens of steerable dishes,
whereas future instruments will be even larger distributed sensor arrays with
a hierarchy of phased array elements. For such arrays to provide meaningful
output (images), accurate calibration is of critical importance. Calibration
must solve for the unknown antenna gains and phases, as well as the unknown
atmospheric and ionospheric disturbances. Future telescopes will have a large
number of elements and a large field of view. In this case the parameters are
strongly direction dependent, resulting in a large number of unknown
parameters even if appropriately constrained physical or phenomenological
descriptions are used. This makes calibration a daunting parameter estimation
task.

\section{Introduction}

\begin{figure}
\centering
\includegraphics[width=\columnwidth]{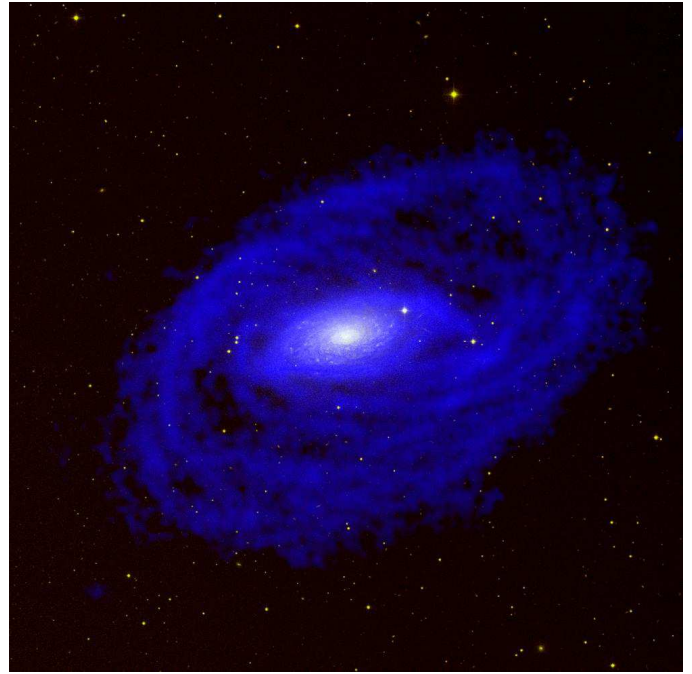}
\caption{Image of the spiral galaxy NGC 5055, showing the structure of the
  neutral hydrogen gas observed with the Westerbork Synthesis Radio Telescope
  (blue) superimposed on an optical image of the same galaxy from the Sloan
  digital sky survey (white) \cite{Battaglia2006-1}.
  \label{fig:NGC5055}}
\end{figure}

Astronomers study the physical phenomena outside the Earth's atmosphere by
observing cosmic particles and electromagnetic waves impinging on the
Earth. Each type of observation provides another perspective on the universe
thereby unraveling some mysteries while raising new questions. Over the years,
astronomy has become a true multi-wavelength science. A nice demonstration is
provided in Fig.\ \ref{fig:NGC5055}. In this image, the neutral hydrogen gas
observed with the Westerbork Synthesis Radio Telescope (WSRT) exhibits an
intricate extended structure that is completely invisible in the optical image
from the Sloan digital sky survey \cite{SDSS}. The radio observations
therefore provide a radically different view on the dynamics of this galaxy.

Images like Fig.\ \ref{fig:NGC5055} are only possible if the instruments used
to observe different parts of the electromagnetic spectrum provide a similar
resolution. This poses quite a challenge since the resolution of any telescope
is determined by the ratio of the wavelength and the telescope diameter.
Consequently, the aperture of radio telescopes has to be 5 to 6 orders of
magnitude larger than that of an optical telescope to provide comparable
resolution, i.e.\ radio telescopes should have an aperture of several hundreds
of kilometers. Although it is not feasible to make a dish of this size, it is
possible to synthesize an aperture by building an interferometer, i.e., an
array.

\begin{figure*}
\parbox[b]{.658\columnwidth}{
\includegraphics[width=.658\columnwidth]{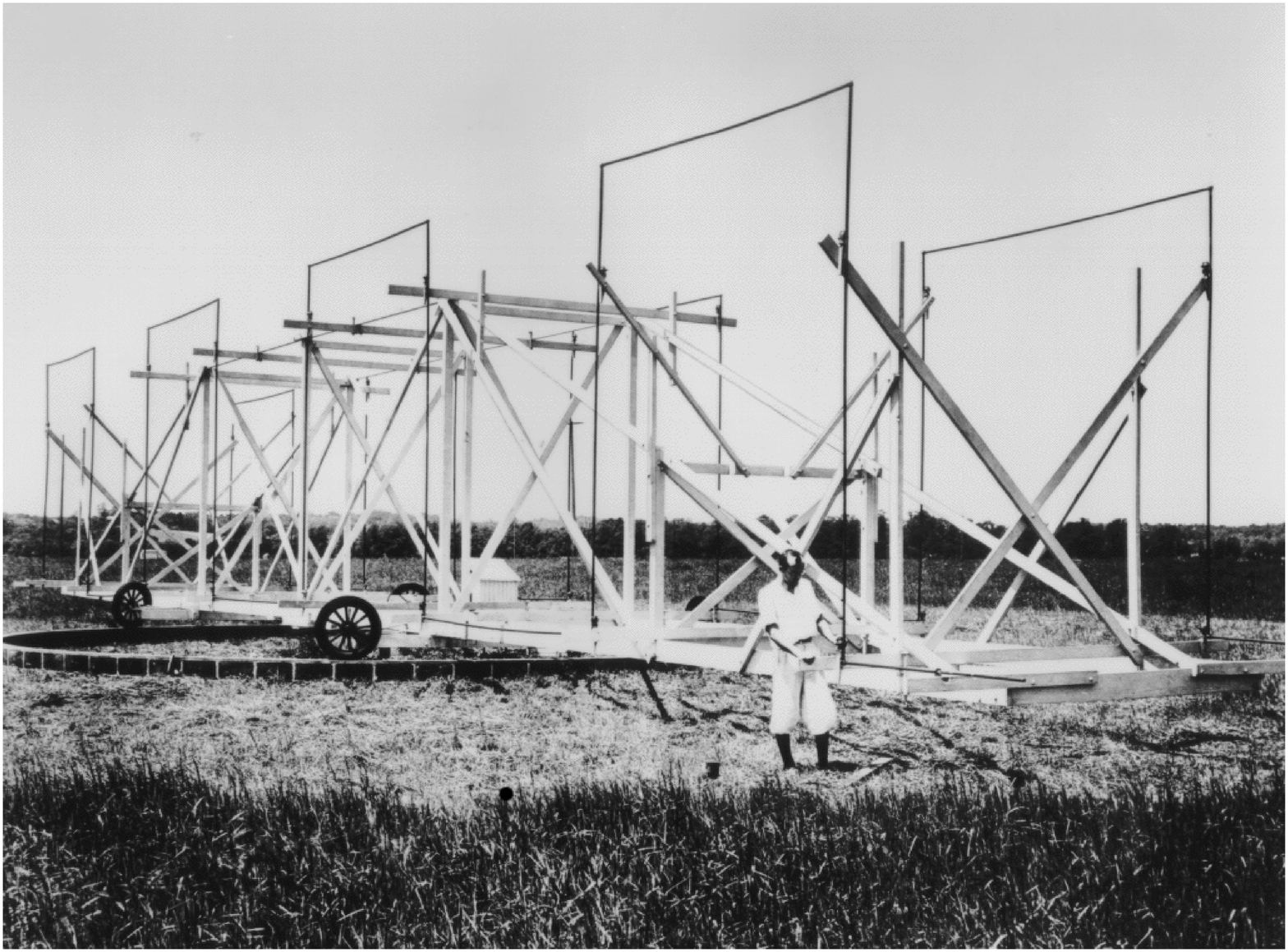}
\includegraphics[width=.658\columnwidth]{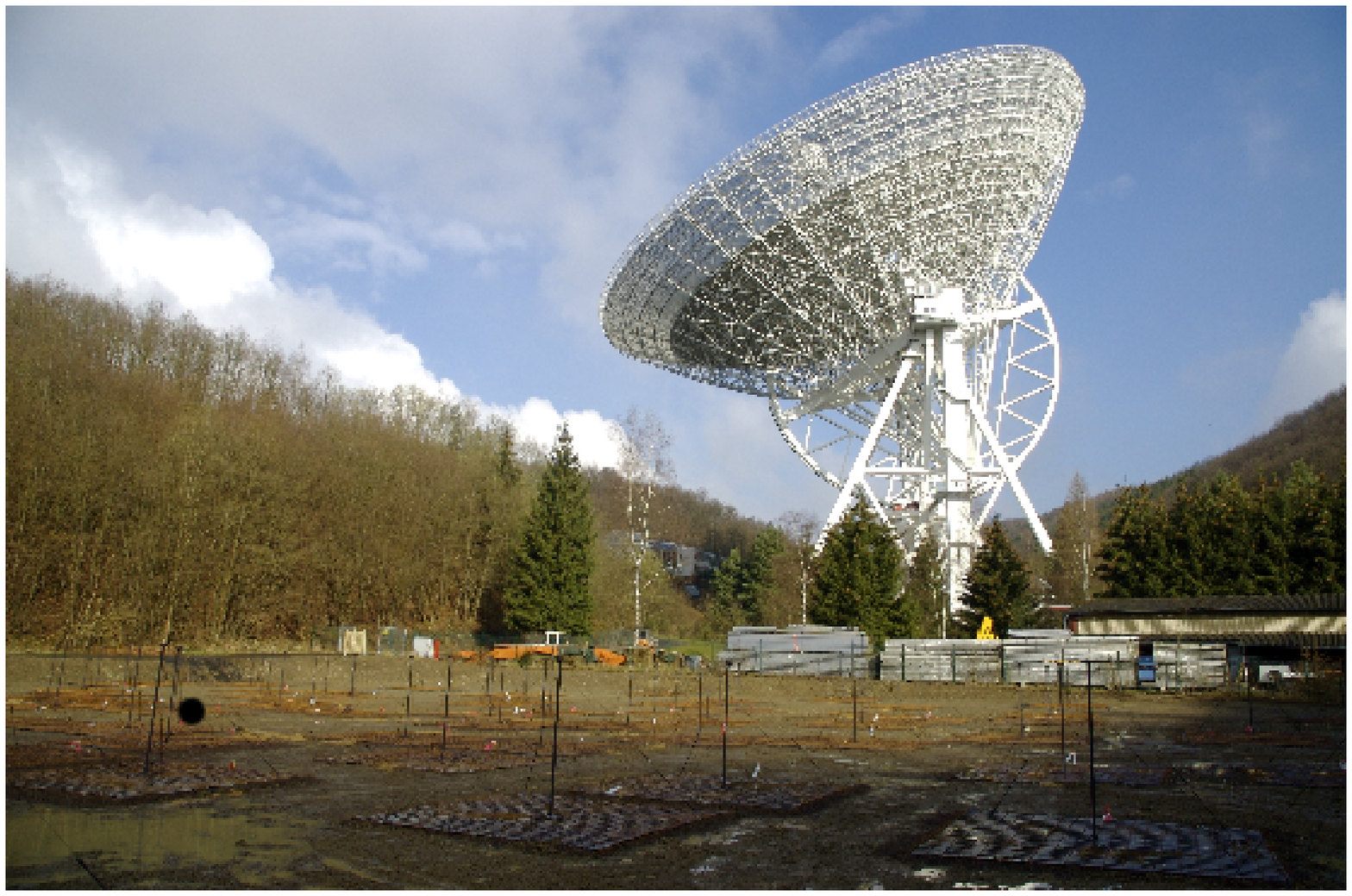}
}
\parbox[b]{.73\columnwidth}{
\includegraphics[width=.73\columnwidth]{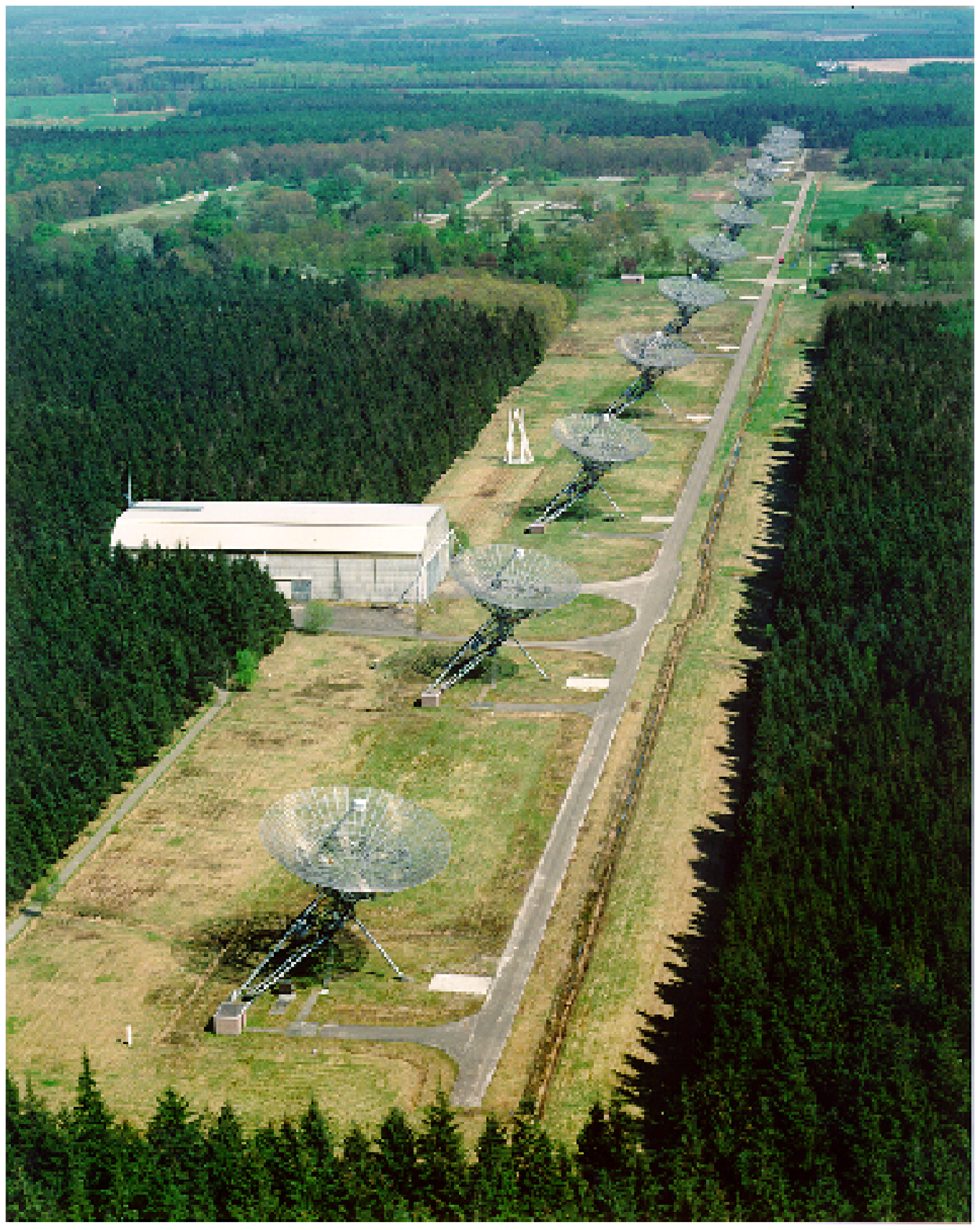}\\[-4.7mm]
}
\parbox[b]{.7\columnwidth}{
    \includegraphics[width=.7\columnwidth]{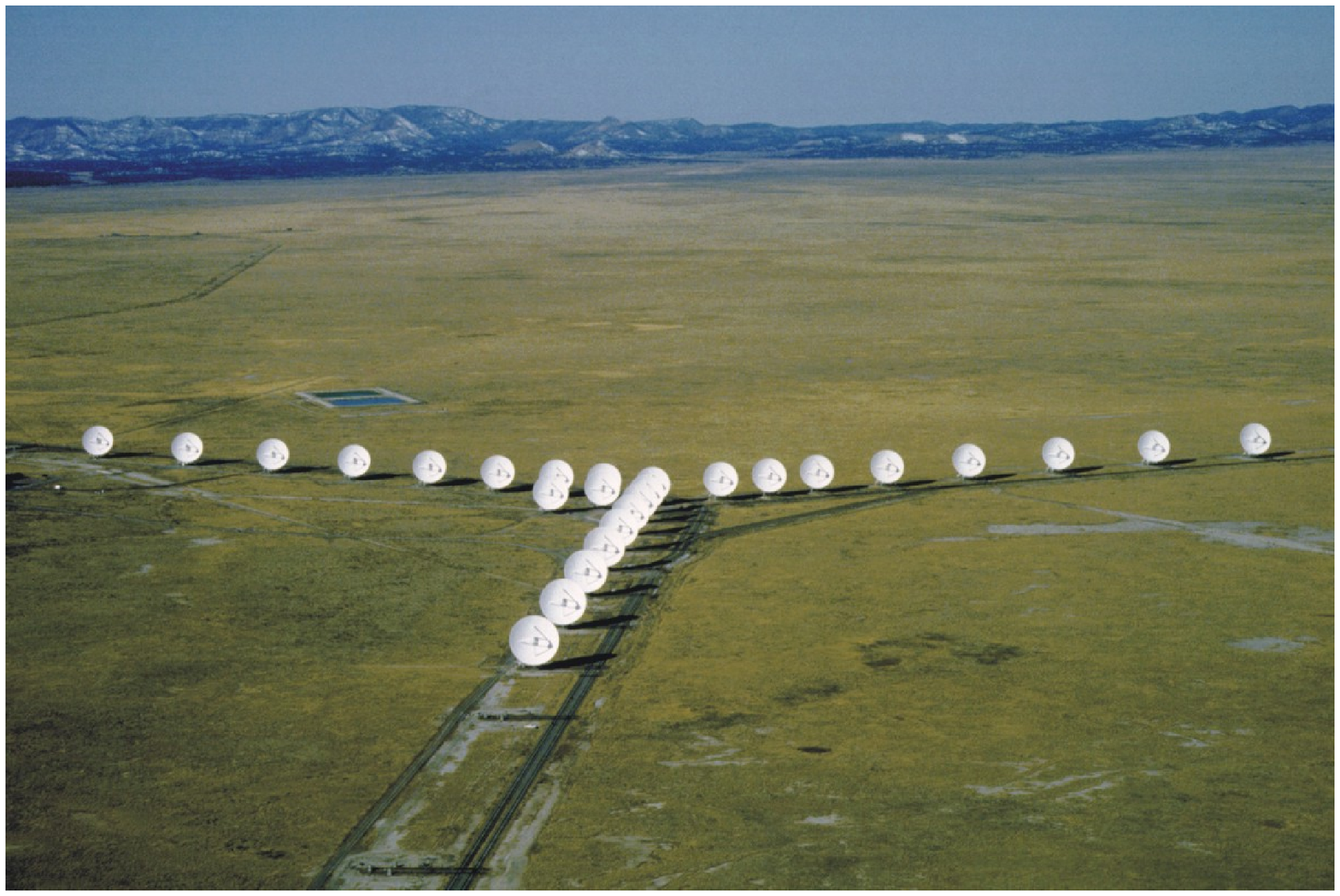}
    \includegraphics[width=.7\columnwidth]{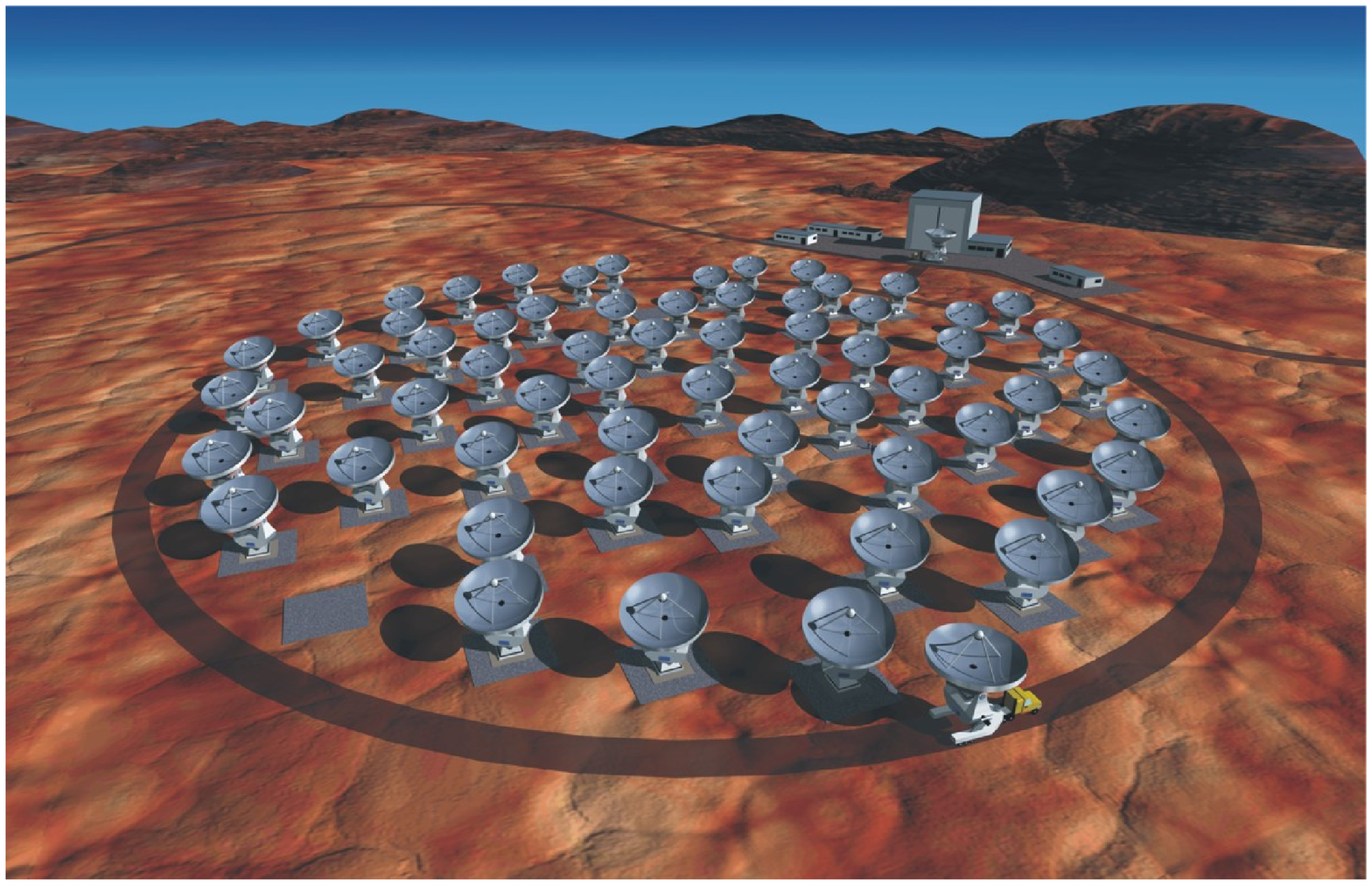}\\[-4.3mm]
}
\caption{The radio telescopes of $(a)$ Jansky, $(b)$ Effelsberg, $(c)$ WSRT,
  $(d)$ VLA, $(e)$ concept for ALMA.}
\label{fig:instruments}
\end{figure*}

Radio astronomy started with the discovery by Karl Jansky, at Bell Telephone
Laboratories in 1928, that the source of unwanted interference in his
short-wave radio transmissions actually came from the Milky Way. For this, he
used the large antenna mounted on a turntable shown in Fig.\
\ref{fig:instruments}$(a)$. Subsequent single-antenna instruments were based
on larger and larger dishes, culminating in the Arecibo telescope (Puerto Rico
1960, 305 m non-steerable dish) and the Effelsberg telescope (Bonn, Germany,
1972, 100 m steerable dish, Fig.\ \ref{fig:instruments}$(b)$). Making larger
steerable dishes is not practical.

An interferometer measures the correlation between two antennas spaced at a
certain distance. Initially used to study a single source passing over the
sky, the principle was used in optical astronomy in the Michelson stellar
interferometer (1890, 1920); the first radio observations using two dipoles
were done by Ryle and Vonberg in 1946 \cite{ryle52}.  Examples of subsequent
instruments are: the Cambridge One Mile Telescope in Cambridge, UK (1964, 2
fixed and one movable 18.3 m dishes); the 3~km WSRT in Westerbork, The
Netherlands (1970, 12 fixed and 2 movable 25 m dishes, Fig.\
\ref{fig:instruments}$(c)$); the 36~km Very Large Array (VLA) in Socorro, New
Mexico, USA (1980, 27 movable 25 m dishes, Fig.\ \ref{fig:instruments}$(d)$);
the 25~km Giant Meter-Wave Radio Telescope (GMRT) in Pune, India (1998, 30
dishes with 45 m diameter). These telescopes use the Earth rotation to obtain
a sequence of correlations for varying antenna baseline orientations relative
to the desired sky image field, resulting in high-resolution images via {\em
  synthesis mapping}. Even larger baselines (up to a few thousand km) were
obtained by combining these instruments into a single instrument using a
technique called VLBI (very long baseline interferometry), where the telescope
outputs are time-stamped and post-processed by correlation at a central
location. An extensive historical overview is presented in
\cite{Thompson2001-1}. In the near future, astronomers are building even
larger arrays, such as the Atacama Large Millimeter Array (ALMA, Chile, 2011,
50 movable 12 m dishes with possible extension to 64 dishes, Fig.\
\ref{fig:instruments}$(e)$), the Low Frequency Array (LOFAR, The Netherlands
(2009, about 30,000 dipole antennas grouped in 36 stations, Fig.\
\ref{fig:hierarchy}), and the Square Kilometer Array (SKA, 2020+, Fig.\
\ref{fig:hierarchy}). A recent issue of the Proceedings of the IEEE (Vol.97,
No.\ 8, Aug.\ 2009) provides overview articles discussing many of the recent
and future telescopes.

High-resolution synthesis imaging would not be possible without accurate
calibration. Initially, the complex antenna gains and phases are unknown; they
have to be estimated. Moreover, propagation through the atmosphere and
ionosphere causes additional phase delays that may create severe
distortions. Finally, image reconstruction or {\em map making} is governed by
finite sample effects: we can only measure correlations on a small set of
baselines. Solving for these three effects is intertwined and creates very
interesting signal processing problems.  In this overview paper, we focus on
the calibration aspects, whereas imaging is covered in a companion paper
\cite{leshem09spm}. The examples provided in this paper are generally borrowed
from low frequency ($<$ 1.5 GHz) instruments, but the framework presented is
applicable to high frequency instruments like ALMA as well.

\section{Interferometry and image formation}
\label{sec:interferometry}

\begin{figure}
\includegraphics[width=\columnwidth]{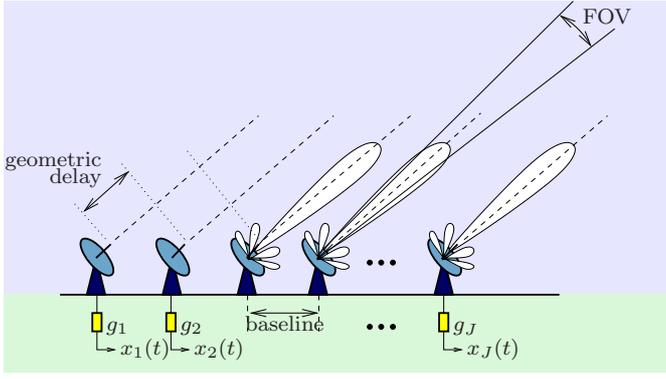}
\caption{Schematic overview of a radio interferometer.}
\label{fig:interferometer}
\end{figure}

The concept of interferometry is illustrated in Fig.\
\ref{fig:interferometer}.  An interferometer measures the spatial coherency of
the incoming electromagnetic field.  This is done by correlating the signals
from the individual receivers with each other.  The correlation of each pair
of receiver outputs provides the amplitude and phase of the spatial coherence
function for the baseline defined by the vector pointing from the first to the
second receiver.  These correlations are called the {\em visibilities}.

\subsection*{Ideal measurement model}

To describe this mathematically, let us assume that there are $J$ array
elements called ``antennas'',\footnote{As discussed in Sec.\
  \ref{sec:architecture}, each element may be a phased array itself!}  pointed
at a field with $Q$ point sources. Stack the sampled antenna signals for the
$k$-th narrowband \cite{Zatman1998-1} frequency channel centered at frequency
$f_k$ into a $J\times 1$ vector $\vx(n)$. For notational convenience, we will
drop the dependence on frequency from the notation in most of the paper. Then
we can model $\vx(n)$ as
\begin{eqnarray}
    \vx(n) & = & \sum_{q=1}^Q \ba_{q}(n) s_{q}(n) + \bn(n)
\label{eq:signalmodel}
\end{eqnarray}
where $s_{q}(n)$ is the signal from the $q$-th source at time sample $n$ and
frequency $f_k$, $\ba_{q}(n)$ is the array response vector for this source,
and $\bn(n)$ is the noise sample vector.  $s_{q}(n)$ and $\bn(n)$ are baseband
complex envelope representations of zero mean wide sense stationary white
Gaussian random processes sampled at the Nyquist rate.

Due to Earth rotation the geometrical delay component of $\ba_{q}(n)$ changes
slowly with time, which is a critical feature exploited in synthesis imaging.
Let $N$ be the number of time samples in a short term integration (STI)
interval.  We assume that $\ba_{q}(n)$ is (relatively) constant over such an
interval, so that, for the $m$-th interval, $\vx(n)$ is wide sense stationary
over $(m-1) N \leq n \leq mN -1$. A single STI autocovariance is defined as
\begin{equation}
    \bR_{m}  =  E \{ \vx (n) \, \vx^H (n) \} 
     =  \bA_{m} \bSigma_s \bA_{m}^H + \bSigma_n , 
\label{eqn:covModel} 
\end{equation} 
    where  $\bR_{m}$ has size $J\times J$,
\begin{eqnarray}
  \bA_{m} & = & \left[\rule{0pt}{2ex} \ba_{1}((m-1)N), \cdots , 
    \ba_{Q}((m-1)N) \right] 
  \nonumber \\
  \bSigma_s & = & \text{diag} \{ [ \sigma^2_{1} , \cdots  
  ,\sigma^2_{Q} ] \} 
  \nonumber \\
  \bSigma_n & = & E \{ \bn(n) \, 
  \bn^H (n) \} 
  \; = \; \text{diag} \{ [ \sigma_{n,1}^2, \cdots , 
  \sigma_{n,J}^2  ] \} . \nonumber
\end{eqnarray} 
Here, $\sigma^2_{q}$ is the variance of the $q$-th source.  Noise is assumed
to be independent but not evenly distributed across the array, and the noise
variances $\sigma_{n,j}^2$ are unknown. With some abuse of notation, the
subscript $n$ in $\bSigma_n$ and $\bsigma_{n,j}$ indicates ``noise''.

Each matrix element of $\left ( \bR_m \right )_{ij}$ represents the
interferometric correlation along the baseline vector between the antennas $i$
and $j$ in the array. The corresponding short term integration sample
covariance estimate is
\begin{equation}
    \hat{\bR}_{m} = \frac{1}{N} \sum_{n=(m-1)N}^{mN-1} \vx(n) \vx^H(n)
    \,, 
    \nonumber
\end{equation}
and this is what the interferometer measures for subsequent processing.  In
practical instruments, the short-term integration interval is often in the
order of 1 to 30 seconds, the total observation can span up to 12 hours, and
the number of frequency bins is highly design-dependent ranging in order of
magnitude from 10 to $10^5$.

Under ideal circumstances, the array response matrix $\bA_{m}$ is equal to a
phase matrix $\bK_{m}$ due entirely to the geometrical delays associated with
the array and source geometry, and accurately known, at least for the
calibration sources. The columns of $\bK_{m}$, denoted by $\vk_{m,q}$ ($q=1,
\cdots, Q$), are often called the ``Fourier kernel'' and are given by
\begin{eqnarray}
    \vk_{m,q}
    & = & \exp \{\jcmplx \frac{2 \pi f_k}{c}  \mZ^T \vp_{m,q} \}  
    \nonumber \\
    \mZ & = & \left [ \bz_1^T, \cdots, \bz_J^T  \right ]^T, \nonumber
\end{eqnarray}
where $c$ is the speed of light, $\bz_j$ is the position column vector for the
$j$-th array element, and $\vp_{m,q}$ is a unit length vector pointing in the
direction of source $q$ during STI snapshot $m$.

\subsection*{Image formation}

Ignoring the additive noise, the {\em measurement equation}
(\ref{eqn:covModel}), in its simplest form, can be written as
\[
    (\bR_m)_{ij}  = 
          \sum_{q=1}^Q \,
            I(\vp_q) \, e^{-\jcmplx\, (\bz_i - \bz_j)^T \vp_{m,q}}  
\]
where $(\bR_m)_{ij}$ is the measured correlation between antennas $i$ and $j$
at STI interval $m$, $I(\cdot)$ is the brightness image {(\em map)} of
interest, and $\vp_q$ is the unit direction vector of the $q$-th source at a
fixed reference time.  It describes the relation between the observed
visibilities and the desired source brightness distribution (intensities), and
it has the form of a Fourier transform; it is known in radio astronomy as the
Van Cittert-Zernike theorem \cite{Perley1994-1, Thompson2001-1}.  Image
formation ({\em map making}) is essentially the inversion of this relation.
Unfortunately, we have only a finite set of observations, therefore we can
only obtain a {\em dirty image}:
\begin{eqnarray*}
    I_D(\vp) 
     & := &  \sum_{i,j,m} \; 
                  (\bR_m)_{ij} \; e^{\jcmplx\, (\bz_i - \bz_j)^T \vp}  
      \\
      & = &   \sum_q \; I(\vp_q) \; B(\vp-\vp_q)
\end{eqnarray*}
where $\bp$ corresponds to a pixel in the image, and where the {\em dirty
  beam}, also referred to as synthesized beam or point spread function (psf),
is given by
\[
      B(\vp-\vp_q) \; := \; 
        \sum_{i,j,m} \;  e^{\jcmplx\, (\bz_i - \bz_j)^T (\vp-\vp_{m,q})}  
	\,.
\]
$I_D(\vp)$ is the desired image convolved with the dirty beam, essentially a
non-ideal point spread function due to the finite sample effect.  Every point
source excites a beam $B(\cdot)$ centered at its location $\vp_q$.  {\em
  Deconvolution} is the process of recovering $I(\cdot)$ from $I_D(\cdot)$
using knowledge of the dirty beam.  A standard algorithm for doing this is
CLEAN \cite{hogbom74}. The autocorrelations are often not used in the image
formation process to reduce the impact of errors in the calibration of the
additive noise on the resulting image. More details are shown in
\cite{Leshem2000-1} and in the companion paper \cite{leshem09spm}.

\begin{figure}
\includegraphics[width=\columnwidth]{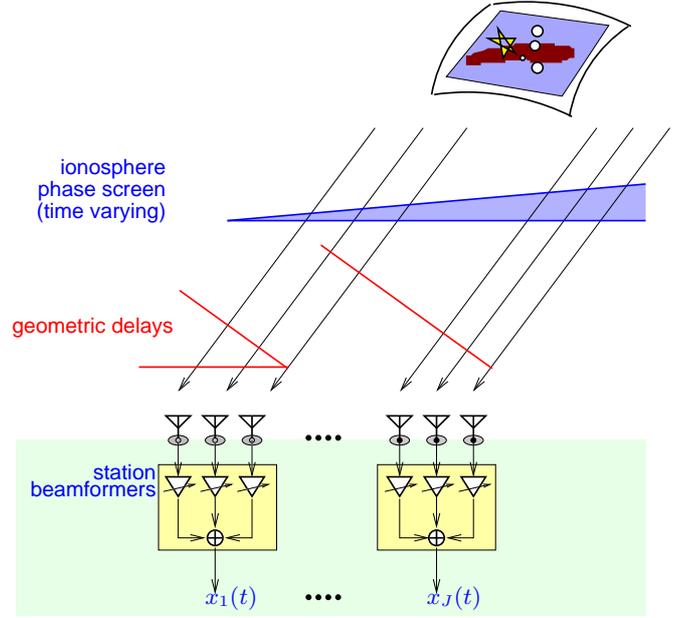}
\caption{A radio interferometer where stations consisting of phased array
    elements replace telescope dishes.  The ionosphere adds phase delays to
    the signal paths. If the ionospheric electron density has the form of a {\em
    wedge}, it will simply shift the apparent positions of all sources.}
\label{fig:interferometer1}
\end{figure}

\subsection*{Non-ideal measurements}

Although the previous equations suggest that it is rather straightforward to
make an image from radio interferometer data, there are
several effects that make matters more complicated:
\begin{itemize}
\item {\em Receiver element complex gain variations.}  Astronomical signals
  are very weak, and radio telescopes therefore need to be very sensitive.
  This sensitivity is inversely proportional to the (thermal)
  noise. This dictates the use of low-noise amplifiers, which are sometimes
  even cryogenically cooled.  Variations in environmental conditions of
  the receiver chain, such as temperature, cause amplitude and
  phase changes in the receiver response. Signals must also be propagated over
  long distances to a central processing facility and, depending on where
  digitization occurs, there can be significant phase and gain variations
  over time along these paths.
\item {\em Instrumental response.} The sensitivity pattern of the individual
  elements, the {\it primary beam}, of an interferometer will never be
  perfect. Although it is steered towards the source of interest, the
  sensitivity in other directions (the {\it side lobe response}) on the sky
  will not be zero. This poses a challenge in the observation of very weak
  sources which may be hampered by signals from strong sources that are
  received via the side lobes, but are still competing with the signal of
  interest. The algorithms correcting for the instrumental response assume
  that the sensitivity pattern is known. This may not be true with the desired
  accuracy if the array is not yet calibrated.
\item {\em Propagation effects.} Ionospheric and tropospheric turbulence cause
  time-varying refraction and diffraction, which has a profound effect on the
  propagation of radio waves. As illustrated in Fig.\
  \ref{fig:interferometer1}, in the simplest cases this leads to a shift in
  the apparent position of the sources.  More generally, this leads to image
  distortions that are comparable to the distortions one sees when looking at
  ceiling lights from the bottom of a swimming pool.
\end{itemize} 

In practice, $\bA_m$ in \eqref{eqn:covModel} is thus corrupted by a complex
gain matrix $\mG_{m}$ which includes both source direction dependent
perturbations and electronic instrumentation gain errors. It is the objective
of calibration to estimate this matrix and track its changes over the duration
of the observation.  Some corrections (e.g., the complex antenna gain
variations) can be applied directly to the measured correlation data, whereas
other corrections (e.g., the propagation conditions) are direction dependent
and are incorporated in the subsequent imaging algorithms.  Very often, the
estimation of the calibration parameters is done in an iterative loop that
acts on the correlation (visibility) data and image data in turn, e.g., the
self-calibration (Self-Cal) algorithm \cite{cornwell81, pearson84} discussed
in more detail later in this paper.

\section{Telescope architectures}
\label{sec:architecture}

\begin{figure*}
\centering
\begin{minipage}[b]{.6\columnwidth}
\centering
\includegraphics[width=\columnwidth]{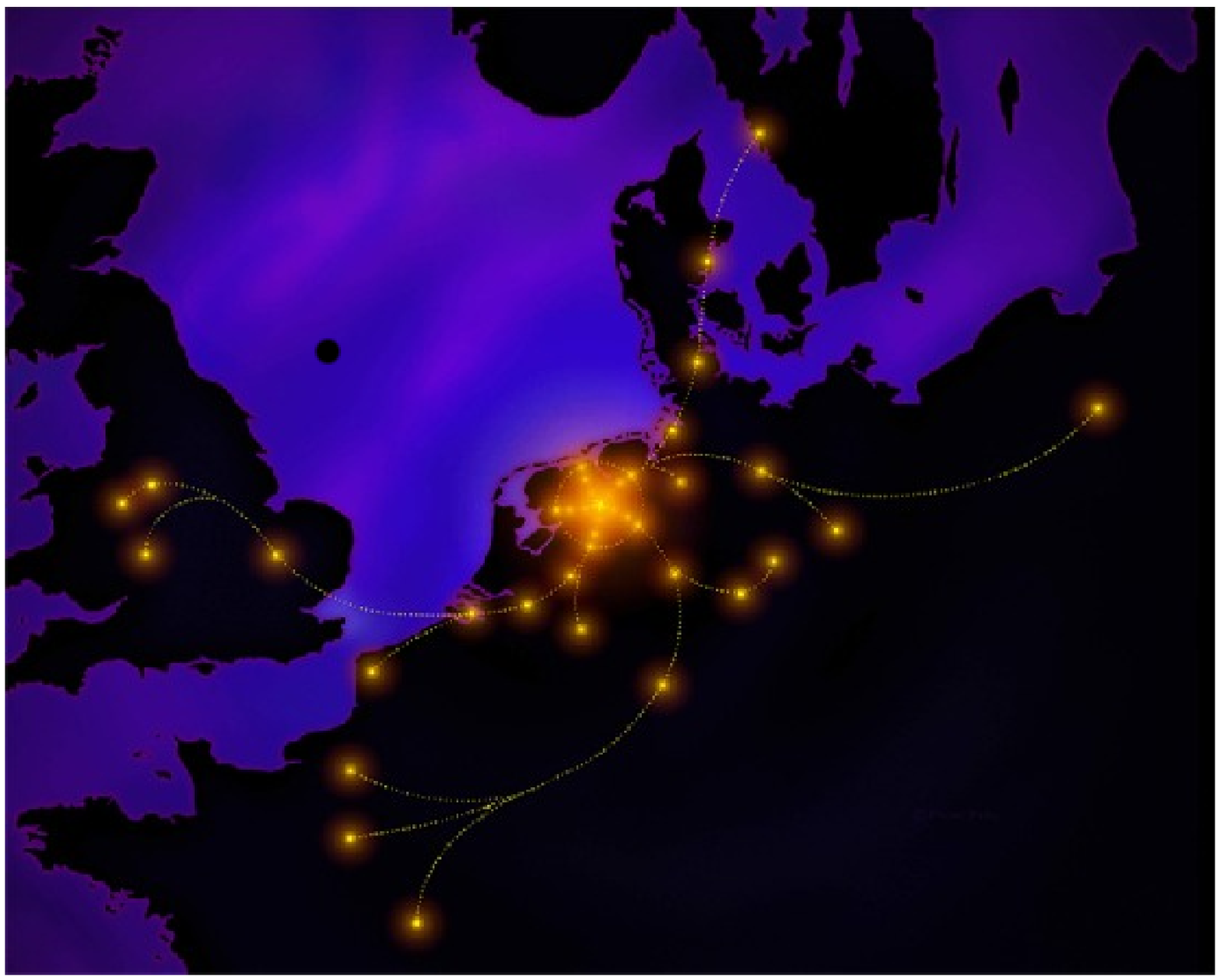}\\[0.5cm]
\includegraphics[width=\columnwidth]{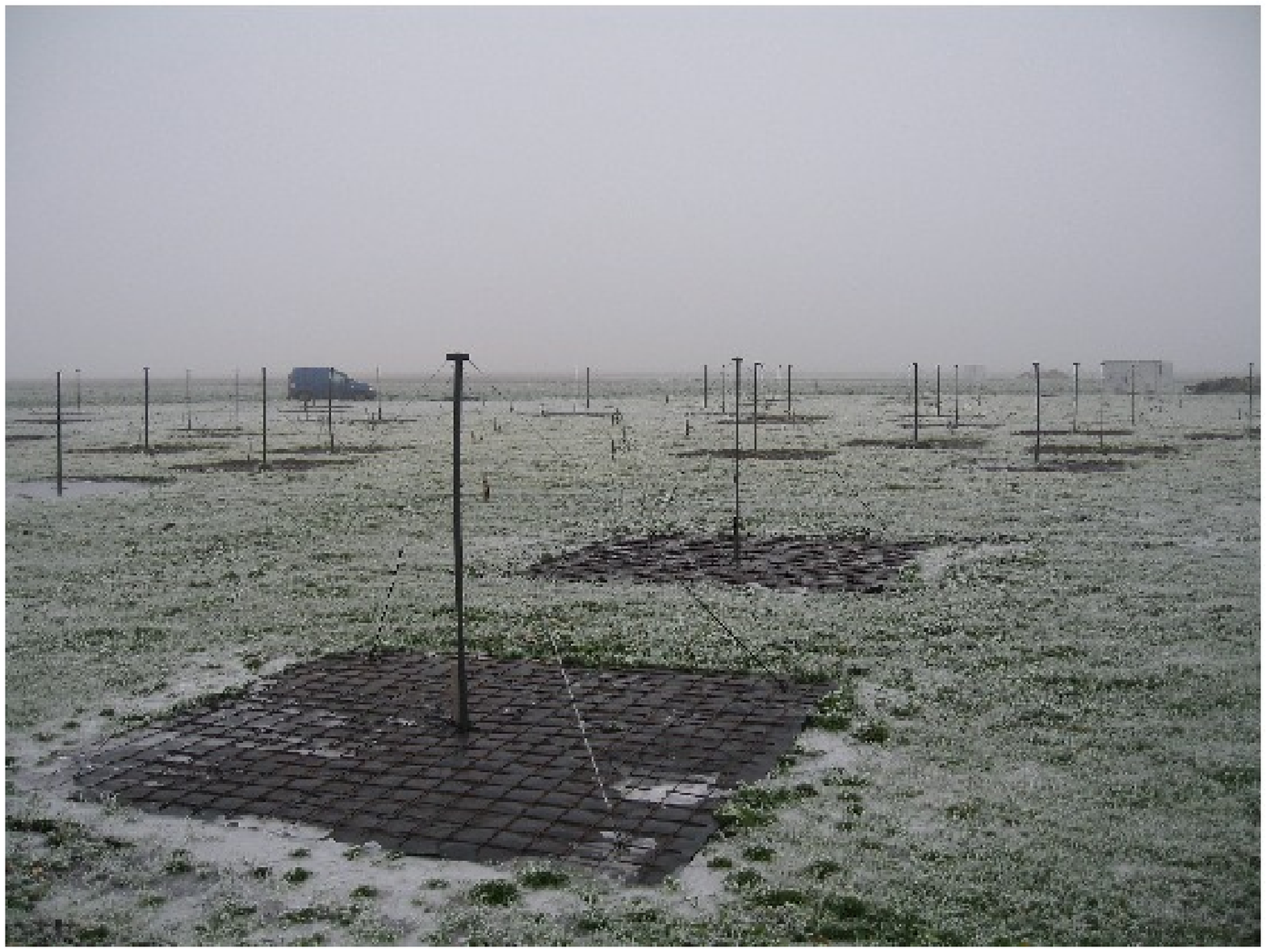}\\[0.5cm]
\includegraphics[width=\columnwidth]{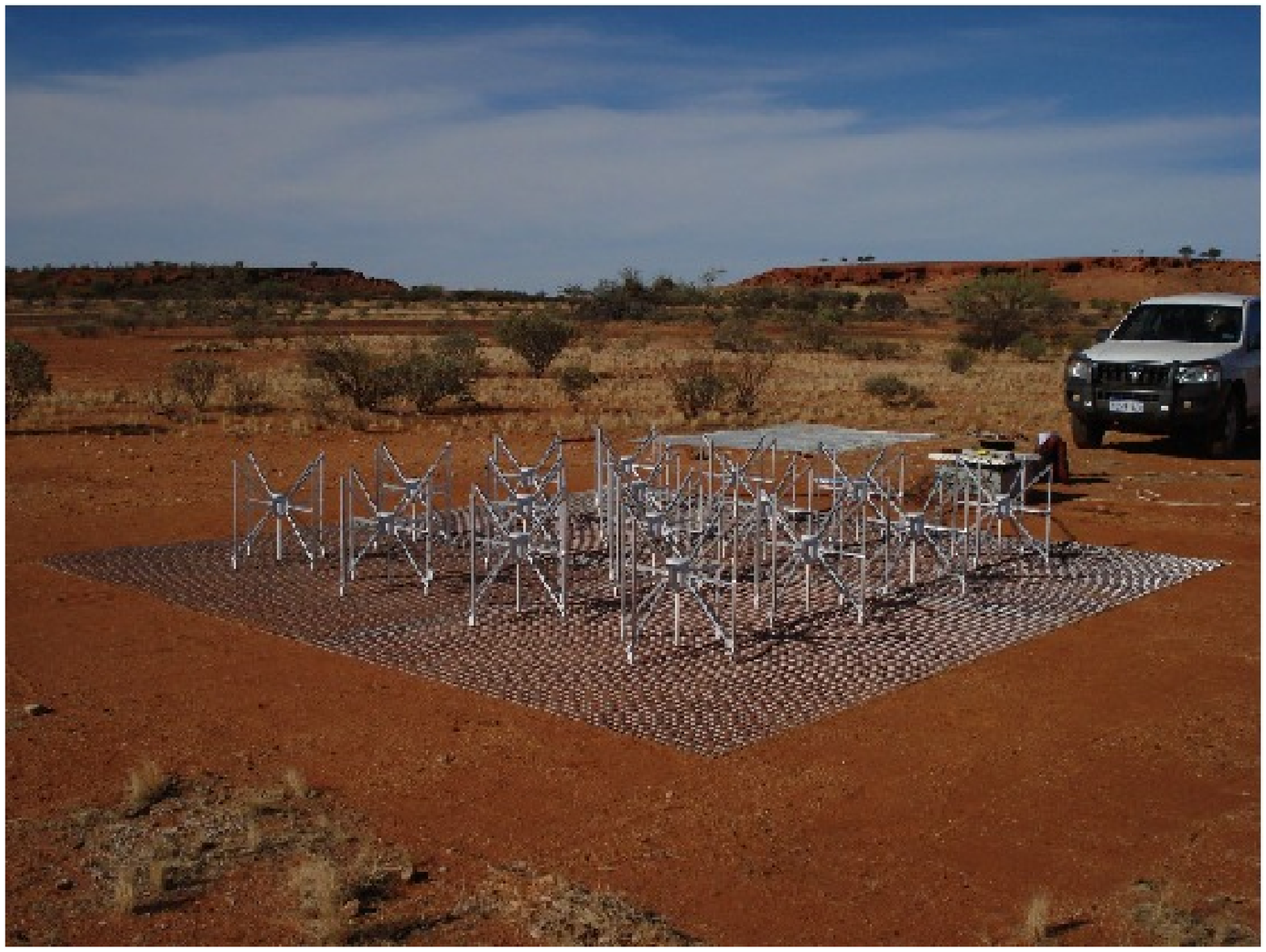}
\end{minipage}
\psfrag{array beam}{array beam}
\psfrag{Array of stations}{array level view}
\psfrag{station}{station level view}
\psfrag{tile}{tile level view}
\psfrag{tile beam}[][]{\shortstack{compound\\ beam}}
\psfrag{antenna beam}[][]{\shortstack{antenna\\ beam}}
\psfrag{station beam}{station beam}
\includegraphics[width=.8\columnwidth]{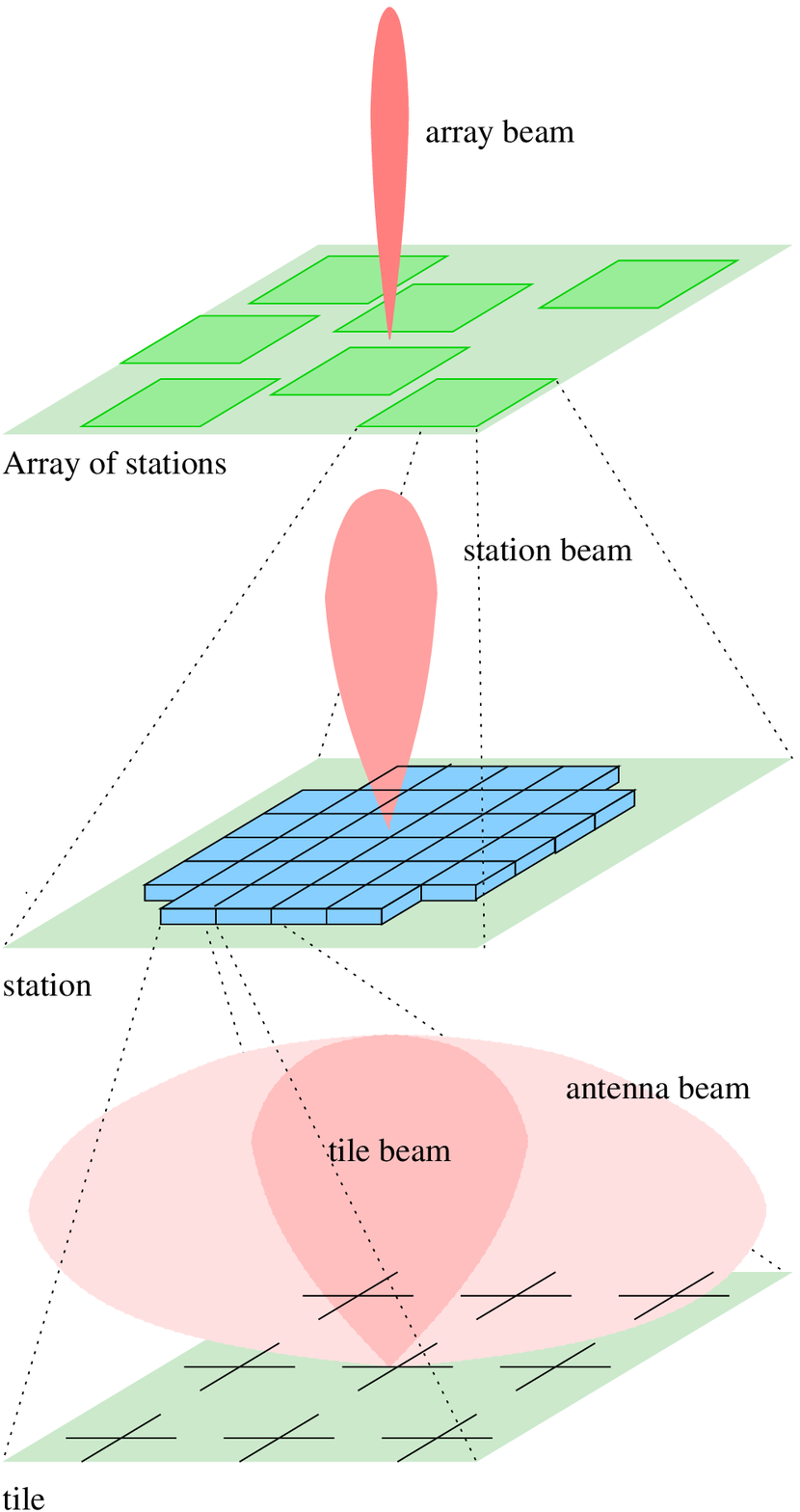}
\begin{minipage}[b]{.6\columnwidth}
\centering
\includegraphics[width=\columnwidth]{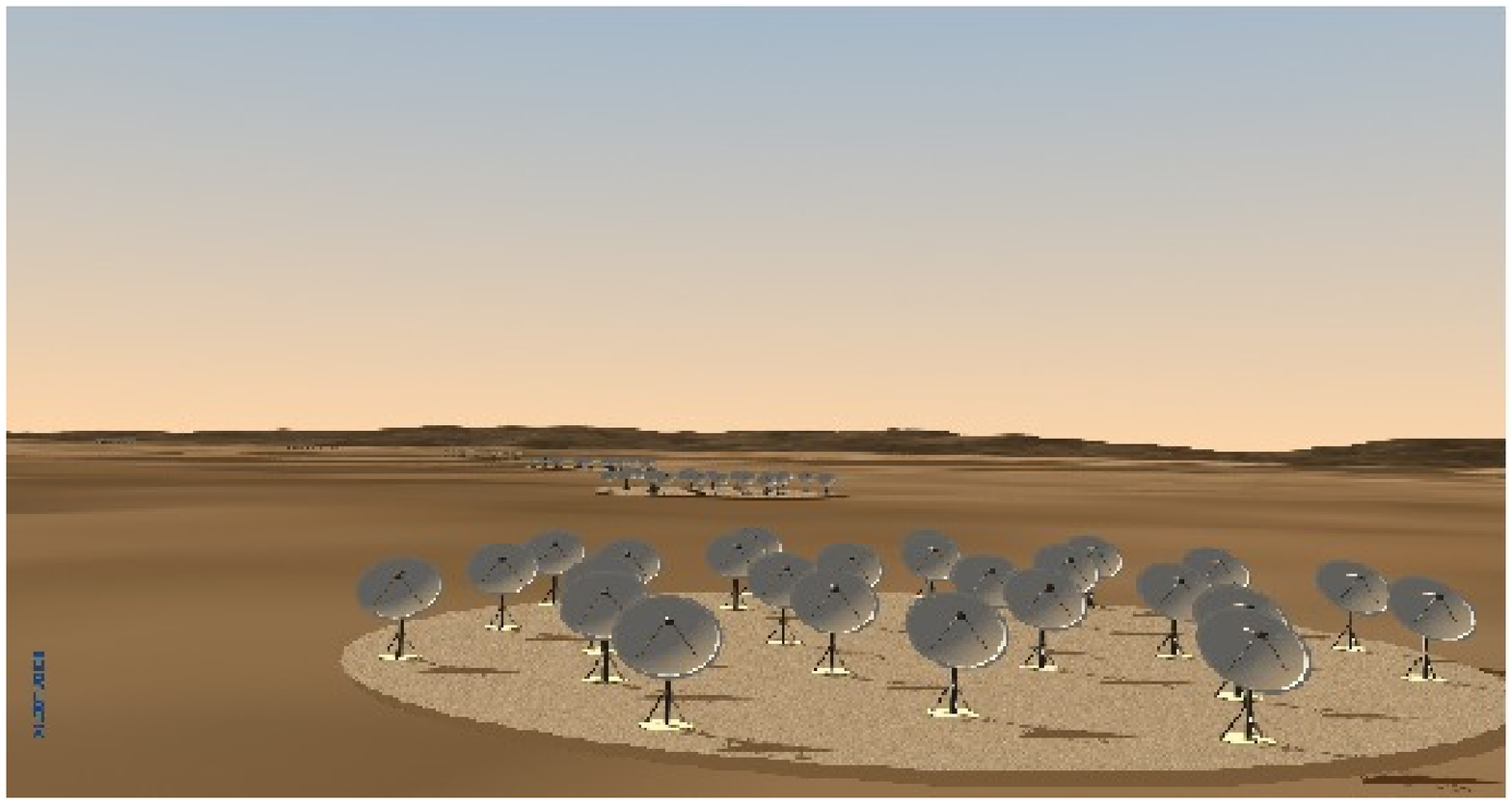}\\[1cm]
\includegraphics[width=\columnwidth]{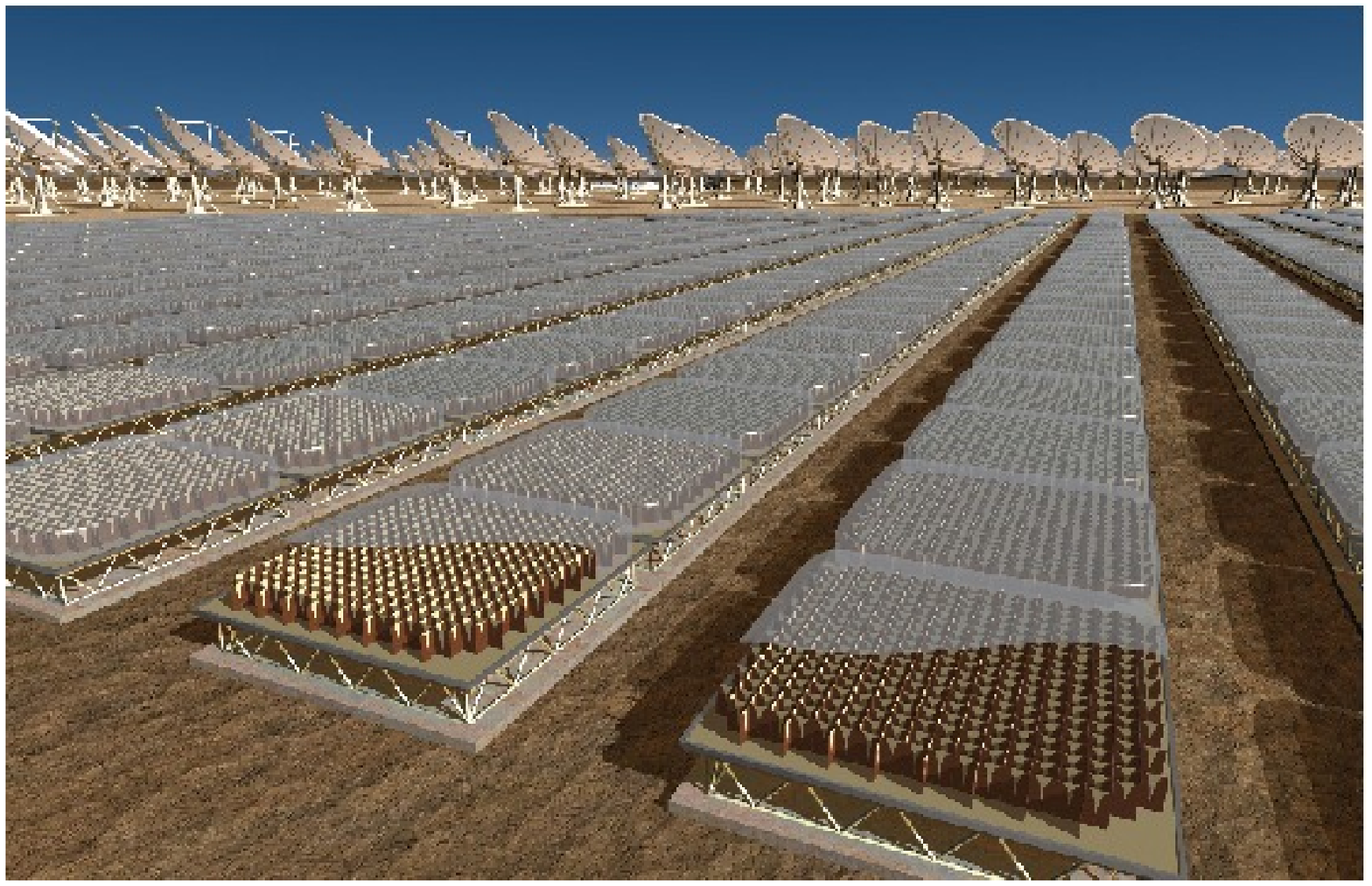}\\[1cm]
\includegraphics[width=\columnwidth]{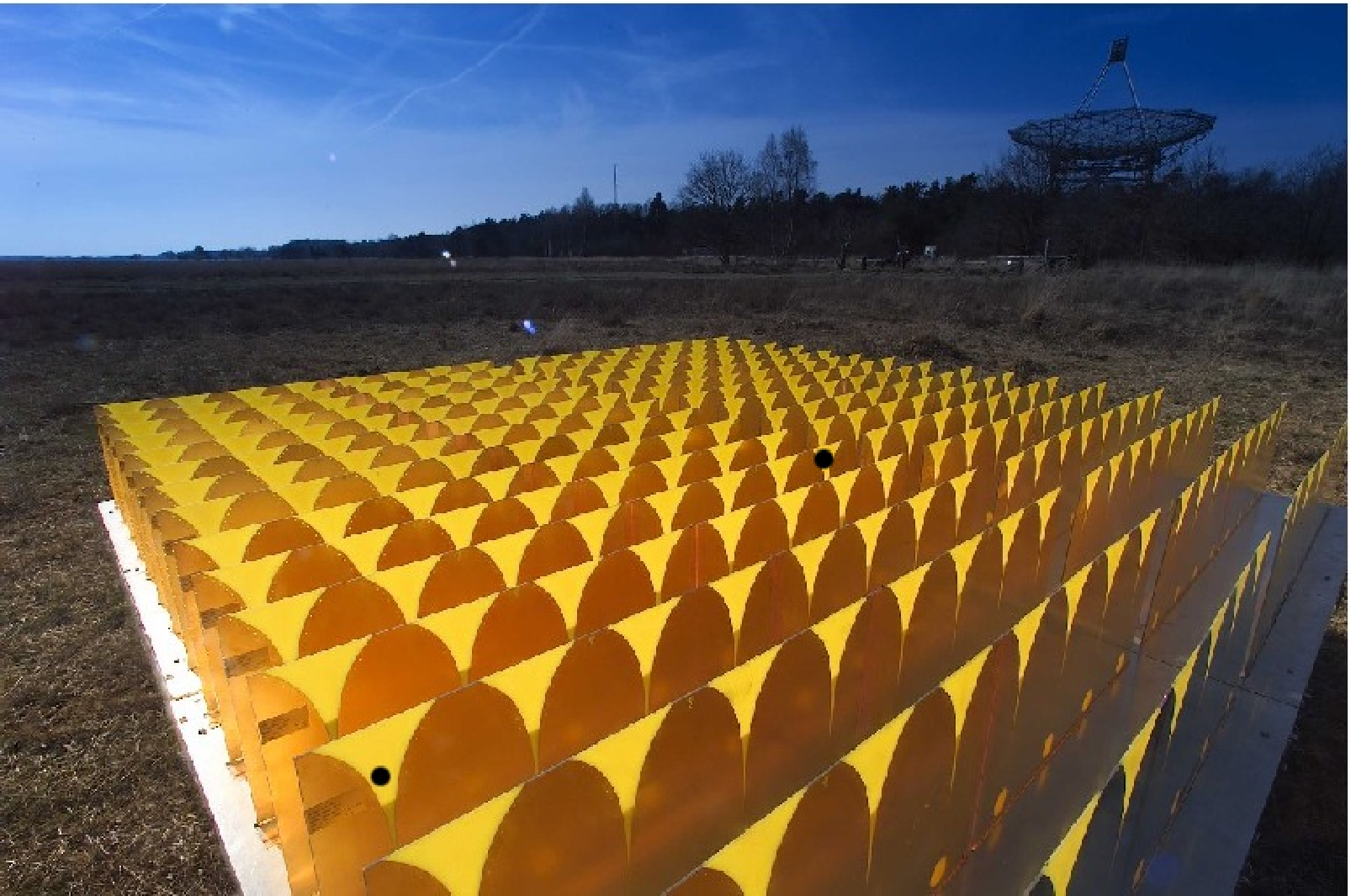}
\end{minipage}
\caption{{\em (Center column)} $(d)$ The beamforming hierarchy with
  the array beam produced by an array of stations at the top and the antenna
  beam at the bottom. Subsequent levels in the hierarchy have beams that
  are narrower and more sensitive. 
  {\em (Left column)} $(a)$ the corresponding concept layout of LOFAR, 
  $(b)$ a LOFAR low band antenna station and $(c)$ a MWA tile.
  {\em (Right column)} $(e)$ a concept for SKA consisting of an array of
  stations, each with small dishes, $(f)$ a concept for the SKA core station, 
  and $(g)$ a SKA demonstrator tile consisting of Vivaldi antennas.
  \label{fig:hierarchy}}
\end{figure*}

The physical model underlying the array calibration depends on the instrument
architecture. This architecture also determines the capabilities of the
telescope and may therefore have a profound effect on the calibration
strategy, as we will see later on.

The Westerbork Synthesis Radio Telescope (WSRT) and the Very Large Array (VLA)
have been the work horses of radio astronomy since the 1970s. Both telescopes
are arrays of 25 m dishes. The size of a dish determines its beamwidth, or
{\em Field of View} (FOV) at a given wavelength, and hence the size of the
resulting image, while the spatial extent of the array determines the
resolution within the FOV. The illumination pattern of the feed on each dish
determines its sensitivity pattern, which is commonly referred to as the
primary beam. These telescopes can also form an instantaneous beam within this
primary by coherent addition of the telescope signals (beamforming). This beam
is called the array beam. Visibilities are measured by correlating the
telescope signals. The baseline vectors on which the visibility function is
observed during a full observation describe a synthesized aperture. The
sampling within this aperture determines the sensitivity pattern of the
synthesis observation, which is referred to as the synthesized beam or point
spread function (psf) and corresponds to the dirty beam in the previous
section. We thus have a beam hierarchy from the primary beam, which has a
relatively large FOV (degrees) and relatively low sensitivity, via the
instantaneously formed array beam to the point spread function, which has a
small FOV (arcseconds) and a high sensitivity.

The WSRT and the VLA have their optimum sensitivity at frequencies of 1 to a
few GHz. At lower frequencies, several things change. There are many more
strong sources (e.g.\ synchrotron emission power is proportional to
wavelength), thus even sources far outside the main beam of the psf may show
their effect due to non-ideal spatial sampling. At low frequencies, the
ionosphere is also much more variable (the phase delays are proportional to
wavelength). Observations at these frequencies are therefore more challenging
and require considerable processing power for proper calibration. High
dynamic range imaging at these frequencies has therefore only recently been
considered.

In the Low Frequency Array (LOFAR) \cite{Bregman2000-1, Vos2009-1}, which is
currently being built in The Netherlands, the dishes are replaced by {\em
  stations}, each consisting of many small antennas distributed over an area
of about $100 \times 100$ meter. Some stations are very closely spaced, others
are placed up to several 100~km away from the core.  A station is a phased
array of receiving elements with its own beamformer.  The stations are steered
electronically instead of mechanically, which allows them to respond quickly
to transient phenomena.  The receiving elements can either be individual
antennas (dipoles), or {\em compound elements} (tiles) consisting of multiple
antennas whose signals are combined using an analog beamformer.  This system
concept introduces two additional levels in the beam hierarchy: the compound
(tile) beam and the station beam.  The complete hierarchy of beam patterns is
illustrated in Fig.\ \ref{fig:hierarchy}.  The Murchison Widefield Array (MWA)
has a similar design and purpose as LOFAR, but is placed in the outback of
Western Australia and has a maximum baseline length of about 3~km.
\cite{Lonsdale2008-1}.

At first sight there is not much difference between the calibration of an
array of stations (or dishes) and the calibration of a station itself, but
there are some subtle differences.  A station only contains short baselines
($\sim100$ m), which implies that it provides a much lower resolution than an
array, while its constituents provide a much wider FOV.  As we will see in the
next section, this implies that the calibration becomes more challenging due
to direction dependent effects.  Another challenging aspect stems from the
enhanced flexibility of electronic beamforming: this may result in a less
stable beam than a beam that is produced mechanically with a reflector
dish. Finally, the output of a station beamformer is insufficient for
calibration: for this purpose the station should also provide the correlation
data among individual elements, even if these are not used at higher levels in
the hierarchy.

A compound element (tile) can also be exploited as {\em focal plane array}
(FPA).  In this case, the array is placed in the focal plane of a dish.  This
allows to optimize the illumination of the dish, as it effectively defines a
spatial taper over the aperture of the dish that can be used to create lower
spatial side lobes.  An FPA can also improve the FOV of the dish by providing
multiple beams (off-axis) on the sky.  In this case, the primary beam is the
sensitivity pattern produced if the dish is illuminated by only a single
antenna of the compound element.  The compound beam is the electronically
formed beam produced by illuminating the dish by the FPA.  The FPA concept is
currently under study in The Netherlands \cite{Cappellen2008-1}, United States
\cite{Warnick2006-1, Jeffs2008-1} and Australia \cite{DeBoer2008-1} as part of
the technology road map towards the Square Kilometer Array (SKA)
\cite{Hall2004-1}.

SKA is a future telescope that is currently in the concept phase.  It is a
wide band instrument that will cover the frequency range from 70 MHz to above
25 GHz.  Apart from cost the design is driven by a trade-off between
sensitivity and survey speed: the speed at which the complete sky can be
observed.  To enable wide band operation, it will probably use a mix of
receiver technologies:
\begin{itemize}
\item {\em Dishes with wide-band single pixel feeds.} At the highest
  frequencies this gives the highest sensitivity. Since this concept provides
  a very stable beam it is well suited for high fidelity imaging.
\item {\em Dishes with focal plane arrays.} At intermediate frequencies FPAs 
    can be used to enlarge the FOV of a single dish in a cost effective way. 
  \item {\em Aperture arrays.} At low frequencies, it is easier to obtain a
    large collecting area, hence sensitivity, by using dipoles instead of
    dishes.  Aperture arrays have the additional advantage of being very
    flexible: by duplicating the receiver chains one can have multiple
    independent beams on the sky, bandwidth can be traded against the number
    of beams on the sky, and electronic beamforming provides a quick response
    time to transient events.
\end{itemize}
The configuration of SKA is still under study. Current concepts include a
dense core that contains, e.g., half of all receivers within a diameter of 5
km, stations consisting of aperture arrays out to a maximal baseline of 180
km, and dishes out to a maximal baseline of 3000 km.

\section{Calibration scenarios}

In the previous section, we introduced several telescope architectures, each
with different characteristics, hierarchy, and parameterizations of the
observed data model.  This will lead to a wide variety of calibration
requirements and approaches.  Fortunately, it is possible to discuss this in a
more structured manner by using only four different scenarios
\cite{Lonsdale2004-1}, each of which can be described by a distinct
specialization of the measurement equation \cite{Hamaker1996-1}. Each scenario
considers the calibration of an array of elements with complex gain variations
and spatially varying propagation effects. The scenarios compare the {\em
  array aperture} (the length of the largest baseline) to the {\em field of
  View} (FOV, the beamwidth of each individual array element) and the {\em
  isoplanatic patch size}, i.e., the scale at which the ionosphere/troposphere
can be considered constant.

\begin{figure*}
\centering
\includegraphics[width=\columnwidth]{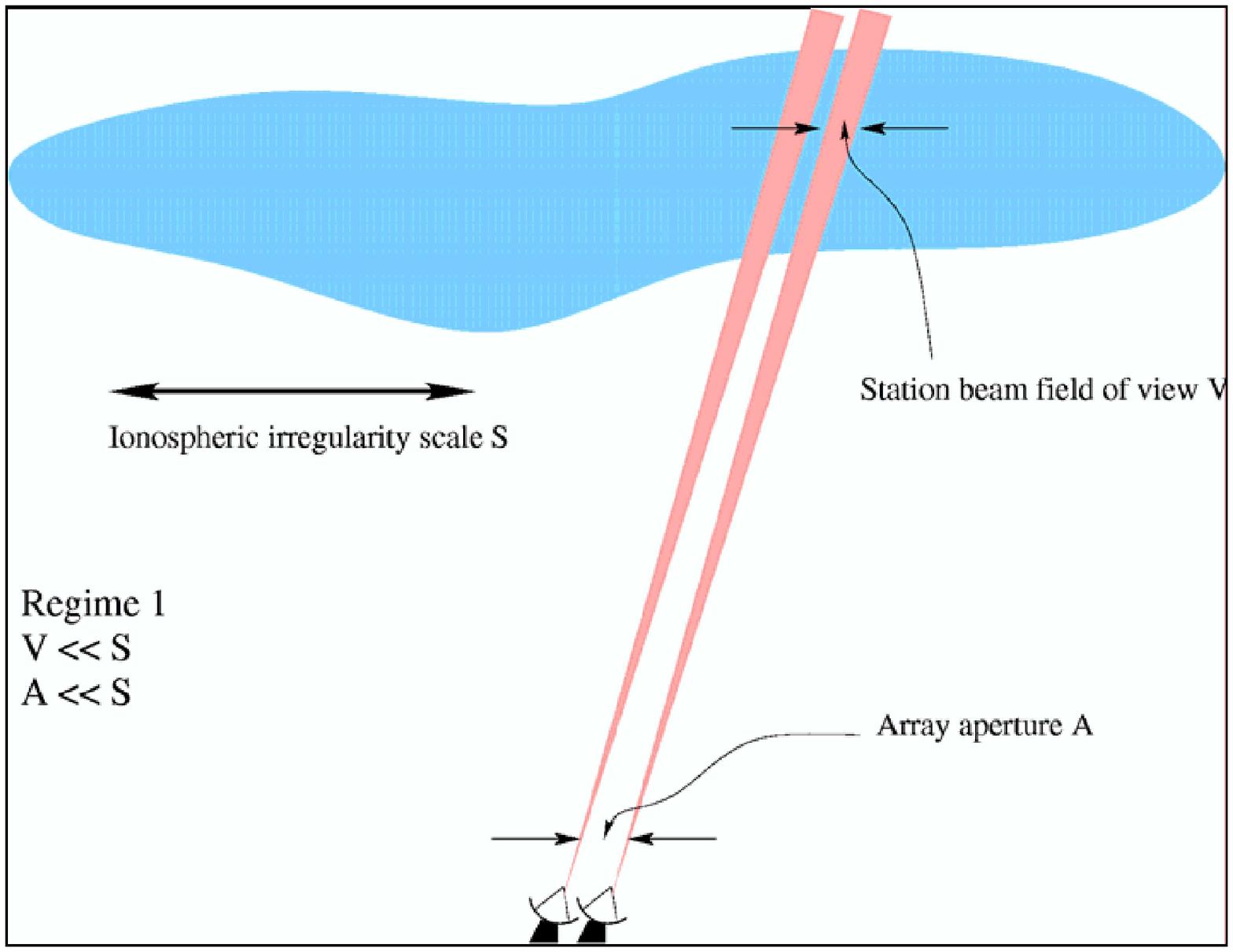}
\includegraphics[width=\columnwidth]{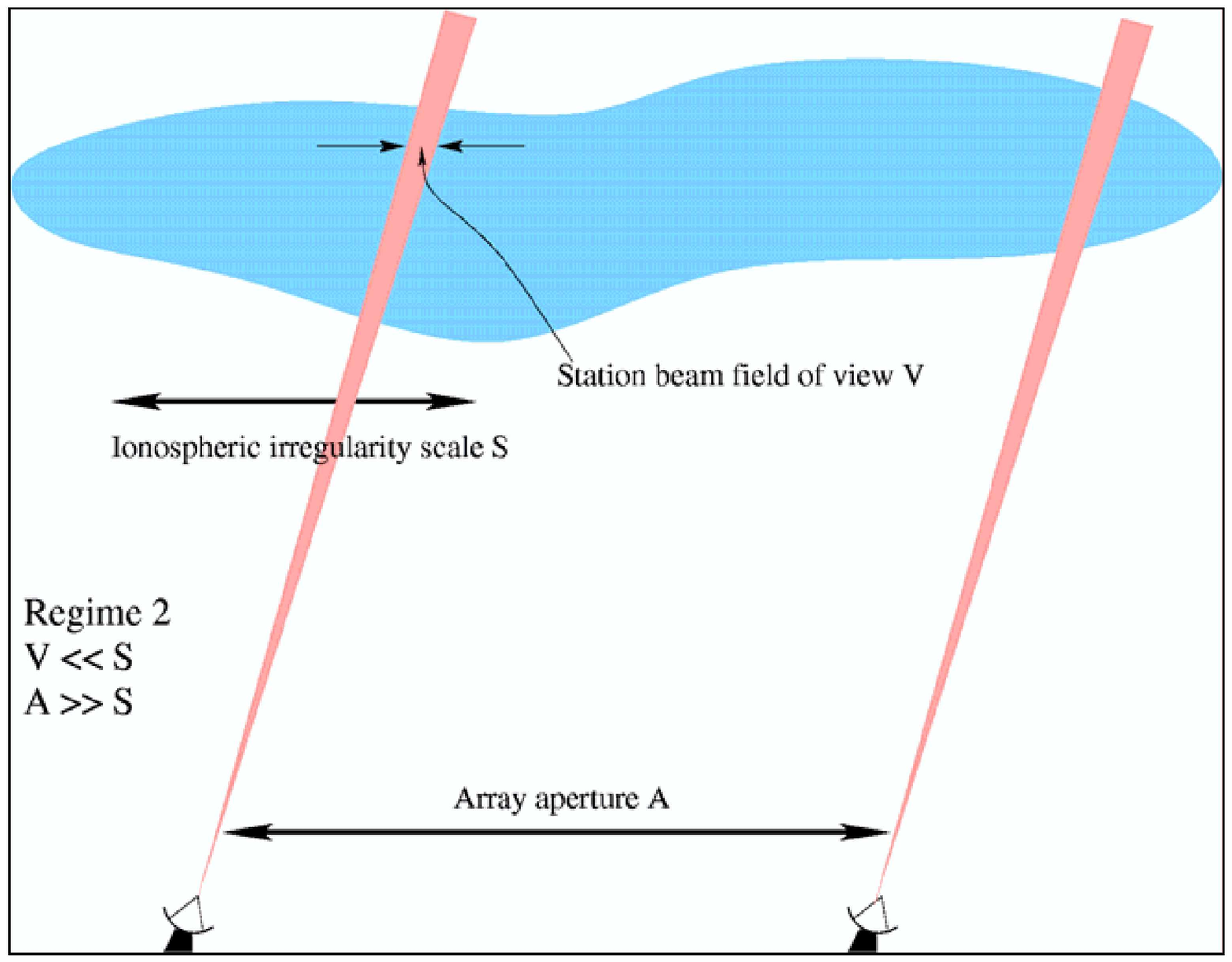}
\includegraphics[width=\columnwidth]{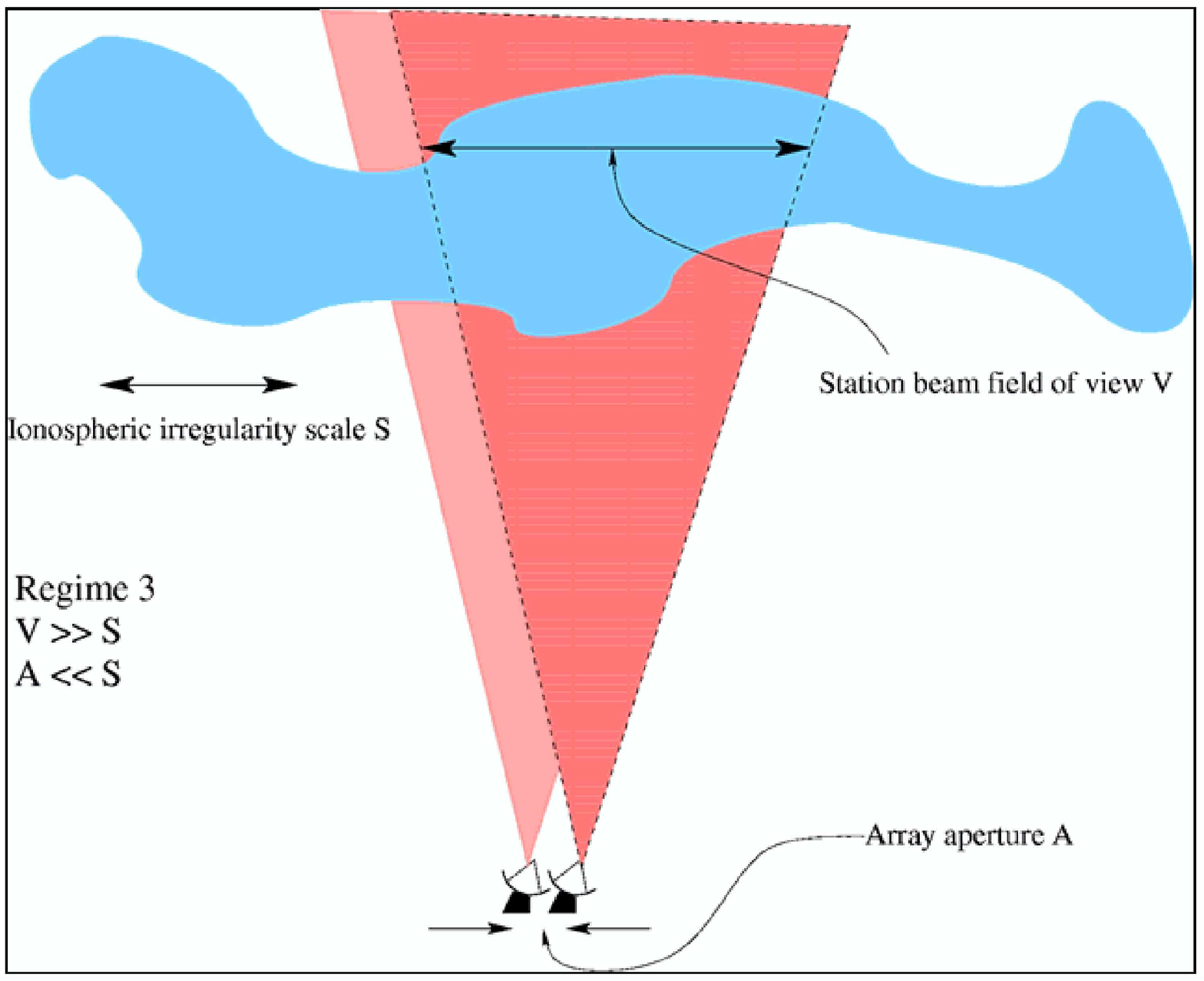}
\includegraphics[width=\columnwidth]{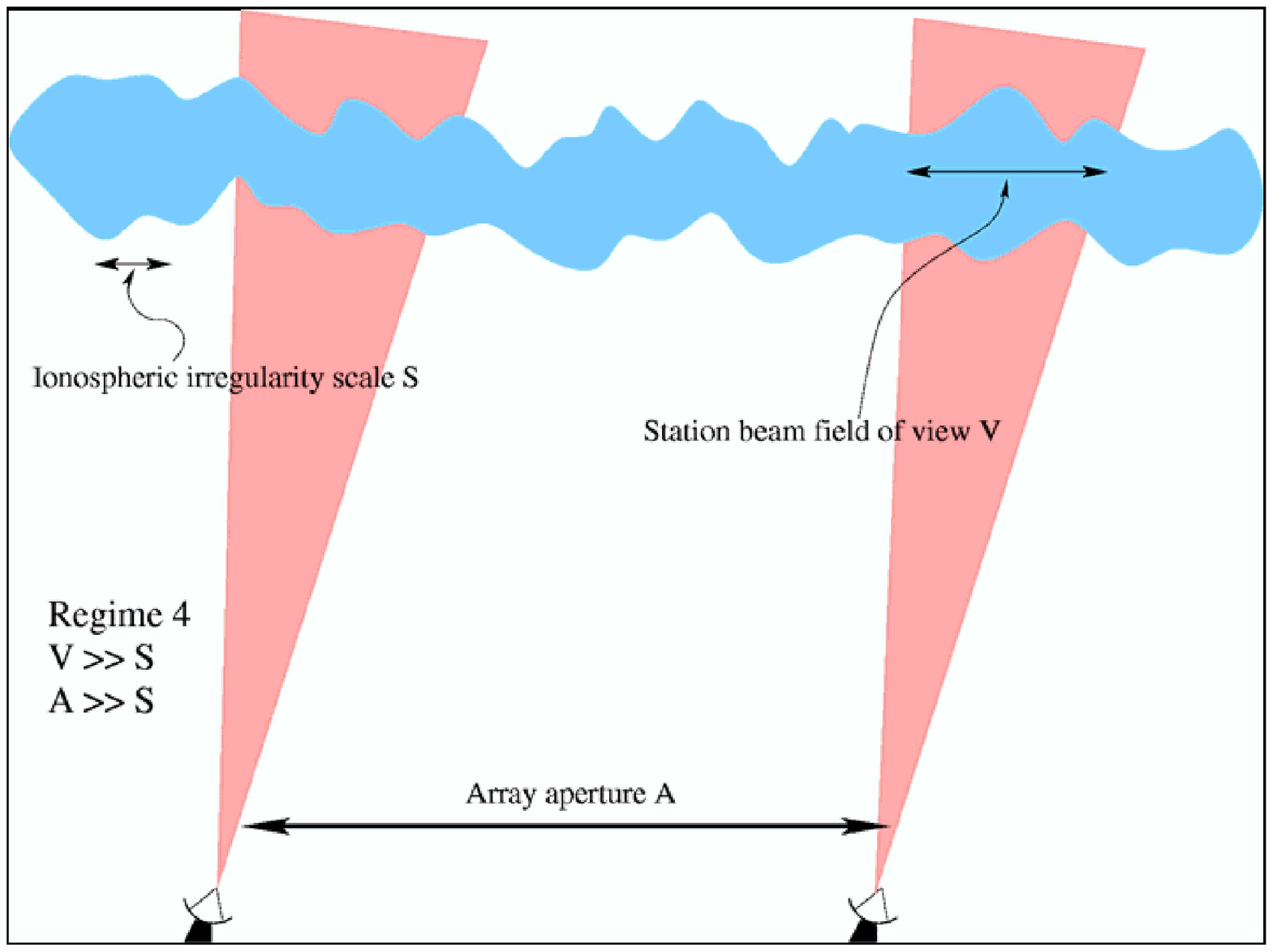}
\caption{Calibration scenarios 1 through 4 (top left to bottom right) as
  defined by Lonsdale \cite{Lonsdale2004-1}}
\label{fig:lonsdale}
\end{figure*}

{\em Scenario 1.} As shown in Fig.\ \ref{fig:lonsdale} (top left), the
receiving elements of the array have a small FOV and the maximum baseline is
short. In this case, all receiving elements and all lines of sight within the
FOV experience the same propagation conditions: the propagation effects do not
distort the image. This scenario represents the case in which direction
dependent effects do not play a significant role. The calibration routine can
therefore focus completely on element based gain effects. Since the FOV is
small, it is often possible to calibrate on a single strong source in the FOV,
especially if the array elements can be steered to a nearby calibration
source. Due to its simplicity, this scenario is often used to obtain a first
order calibration for new instruments.

{\em Scenario 2.} In this scenario (Fig.\ \ref{fig:lonsdale}, top right), we
have a large array consisting of elements with a small FOV. Lines of sight
from different elements towards the region of interest are subject to
different propagation conditions, but the propagation conditions for all
lines-of-sight within the FOV of an individual element are the same. The
propagation effects can therefore be merged with the unknown receiver gains of
each element, and the array can be calibrated under the same assumptions as in
the first scenario.  This scenario is valid for most of the interferometers
built in the 1970s and 1980s, such as the WSRT and the VLA, and for VLBI
observations.

{\em Scenario 3.} Fig.\ \ref{fig:lonsdale} (bottom left) depicts the third
scenario in which the elements have a large FOV, but the array is small. This
implies that all lines of sight go through the same propagation path, but that
there may be considerable differences in propagation conditions towards
distinct sources within the FOV. The ionosphere and troposphere thus impose a
direction dependent gain effect that is the same for all elements. This
scenario can also handle instrumental effects that are the same for all
elements (e.g., irregular antenna beamshapes) and is therefore well suited for
the situation of a compact array of identical elements such as the MWA
and a single LOFAR or SKA station.

{\em Scenario 4.} As shown in the bottom right panel of Fig.\
\ref{fig:lonsdale}, the elements have a large FOV and the array has a number
of long baselines. The lines of sight towards each source may experience
propagation conditions that differ for different elements in the array. This
implies that distinct complex gain corrections may be required for each source
and each receiving element.  Calibration is not possible without further
assumptions on stationarity over space, time and/or frequency.  This is the
most general scenario, and valid for future telescopes such as LOFAR, SKA and
ALMA.

\begin{figure}
\framebox{
\parbox[t]{.95\columnwidth}{
   \subsection*{Notation}
   \begin{tabular}{ll}
    $\otimes$ & Kronecker product \\
    $\circ$ & Khatri-Rao (columnwise Kronecker) product \\
    $\odot$ & Schur (entry-wise) matrix product \\
    $\oslash$ & entry-wise matrix division\\
    $( \cdot )^T$ & transpose operator\\
    $( \cdot )^H$ & Hermitian (conjugate) transpose operator\\
    $( \bA )^{\odot \beta}$ & entry-wise power of a matrix \\
    $\vect( \cdot )$ & stacking of the columns of a matrix \\
    $\diag( \cdot )$ & diagonal matrix constructed from a vector \\
    $\overline{a}$ & complex conjugate of $a$ \\
    $\dagger$ & Moore-Penrose pseudo-inverse
   \end{tabular}
}
}
\end{figure}

\section{Array calibration}
\subsection*{Scenario 1}

In scenario 1, the field of view (FOV) of each array element (dish) is small
and it is reasonable to assume that there is only a single calibrator source
within the beam.  Often, the beam will even have to slightly point away from
the field of interest to ``catch'' a nearby strong calibrator source.  The
calibrator should be unresolved, i.e., appear as a point source, as opposed to
the extended structure visible in Fig.\ \ref{fig:NGC5055}.

We will assume that the STI sample covariance matrices $\widehat{\bR}_m$ are
calibrated independently and omit the subscript $m$ for notational
convenience. Continuity over many STIs needs to be exploited under scenario 4
and will thus be discussed later. The data model (measurement equation) under
scenario 1 is given by
\begin{equation}
       \bR = \bG \bK\bSigma_s\bK^H \bG^H \;+\; \bSigma_n \label{eq:scenario1}
\end{equation}
where $\bG = \diag(\bg)$ is a diagonal matrix. For a single calibrator source,
$\bK$ has only a single column representing the geometric phase delays of the
array towards the source, and $\bSigma_s = \sigma_s^2$ is a scalar with the
source power.  Both the direction of the source and its power are known from
tables. Thus, in essence the problem simplifies to
\[
   \bR \;=\; \bg\bg^H \;+\; \bSigma_n
\]
This is recognized as a ``rank-1 factor analysis'' model in multivariate
analysis theory \cite{mardia79,lawley71}.  Given $\widehat{\bR}$, we can solve
for $\bg$ and $\bSigma_n$ in several ways \cite{boonstra02calibration,
  Boonstra2005-1, Wijnholds2006-1}. E.g., any submatrix away from the diagonal
is only dependent on $\bg$ and is rank 1: this allows direct estimation of
$\bg$. In the more general case described by \eqref{eq:scenario1}, multiple
calibrators may be simultaneously present. The required multi-source
calibration is discussed in, e.q., \cite{Pierre1991-1, Fuhrmann1994-1,
  Flanagan2001-1, Wijnholds2009-1}.

\subsection*{Scenario 2}

In this scenario, the ionospheric or tropospheric phases are different for
each array element, but the field of view is narrow, and it is possible to
assign the unknown ionospheric or tropospheric phases to the individual
antennas. Thus, the problem reduces to that of Scenario 1, and the same
calibration solution applies.

\begin{figure*}
\framebox{

\begin{minipage}[t]{.95\columnwidth}
\setlength{\parindent}{1em}
\subsection*{Ionospheric calibration based on a statistical model}

Assume for simplicity a single calibration source at zenith.
The data model is
\[
        \bR = \ba\ba^H \sigma_s^2 + \sigma_n^2 \bI 
\] 
$\ba$ is the spatial signature of the source at frequency $f_k$, as caused
(only) by the ionosphere, and given by
\[
        \ba = \exp(\jcmplx \vphi) \,,\qquad
        \vphi = C \vgt f_k^{-1} 
\]
where $\vphi$ is a vector with $J$ entries representing the ionospheric phases
at each station, vector $\vgt$ contains the Total Electron Content (TEC) seen
by each station, and $C$ is a constant.  The TEC is the integral of the
electron density along the line of sight, and is directly related to a
propagation delay.

The ionosphere is often modeled as a turbulent slab of diffracting medium.
Assuming a single layer and a pure Kolmogorov turbulence process, the
covariance for $\vgt$ is modeled by a power law of the form
\[
    \bC_\tau = \bI - \alpha ( \bD )^{\odot \beta}
\]
with unknown parameters $\alpha$ and $\beta$ (theoretically, $\beta = 5/3$ but
measured values show that deviations are possible).  The matrix $\bD$ contains
the pairwise distances between all antennas, and is known.

Write the model as
\[
    \vect(\bR) = (\bac\otimes\ba)\sigma_s^2 + \vect(\bI) \sigma_n^2 
\]
The observed covariance matrix is $\vect(\bRh) = \vect(\bR(\vgt)) + \bw$,
where the observation noise $\bw$ has covariance $\mC_{\vw} =
\frac{1}{N}(\bR^T \otimes \bR)$.  At this point, we could estimate $\vgt$ by a
Least Squares model matching.  However, we have a priori knowledge on the
parameters $\vgt$, i.e., the covariance $\bC_\tau$.  The maximum a posteriori
(MAP) estimator exploits this knowledge, and leads to
\[
	\hat{\vgt} 
	 =  \argmin_{\vgt}
	       \|\mC_{\vw}^{-\frac{1}{2}}\,\vect(\bRh-\bR(\vgt))\|^2
	       \;+\;
	       \|\mC_{\tau}^{-\frac{1}{2}}\vgt\|^2 \,.
\]
This is solved as a nonlinear least squares problem ($\alpha$ and $\beta$ are
estimated as well).

\end{minipage}
\hspace{1em}
\begin{minipage}[t]{.95\columnwidth}
  The dimensionality can be reduced by introducing $\vgt = \mU \btheta$, where
  $\bU$ contains a reduced set of basis vectors.  These can be
\begin{itemize}
\item Data independent, e.g., simple polynomials, or Zernike polynomials
  \cite{Cotton2002},
\item Data dependent (Karhunen-Loeve), based on an eigenvalue decomposition of
  $\bC_\tau$: $\bC_\tau \approx \bU \bLambda \bU^H$.  Only the dominant
  eigenvectors are retained.
\end{itemize}
Selection of the correct model order is often a tradeoff between reduced
modeling error and increased estimation variance.  Indeed, the simulation
below indicates that the Least Squares estimator (with either a Zernike basis
or a Karhunen-Loeve basis) has an optimal model order, beyond which the mean
squared estimation error increases. The MAP estimator adds an additional term
to the cost function that penalizes ``weak'' parameters and makes it robust to
overmodeling.  For more details, see \cite{tol07isscs}.  The validity of the
turbulence model is experimentally tested on VLA survey data in
\cite{Cohen:2009,Tol:2009}.

   \begin{center}
     \includegraphics[width=\columnwidth]{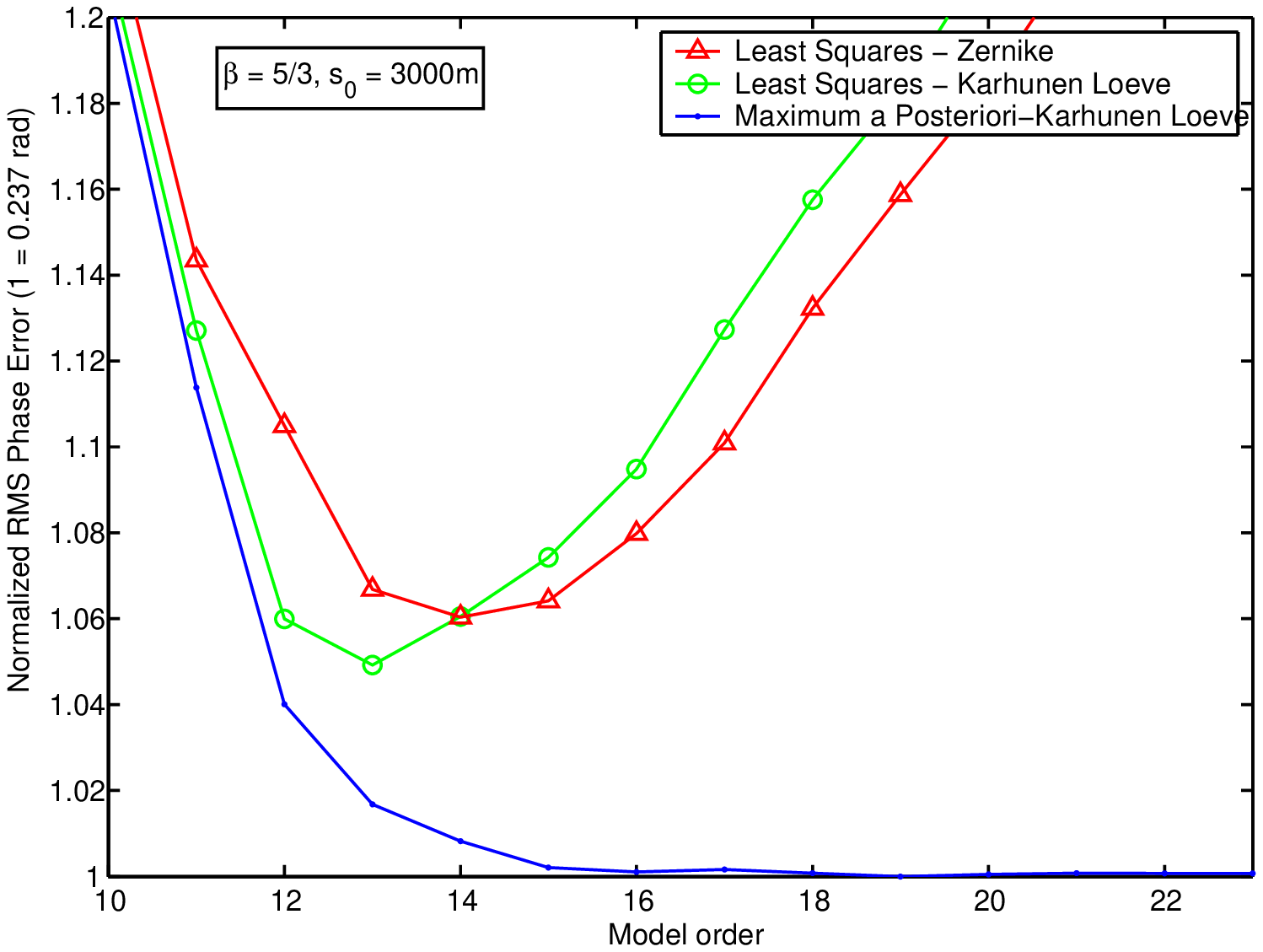}
   \end{center}
   {\small\em Estimation performance as function of model order selection}
\end{minipage}
}
\end{figure*}


\subsection*{Scenario 3}

This scenario is relevant for sufficiently compact arrays, e.g., the
calibration of a SKA or LOFAR station or an array with a relatively small
physical extend like the MWA. The phase and gain of the station beam FOV is
direction dependent but the array elements see the same ionosphere. It is
possible to make a coherent image, but sources may have shifted to different
locations.

Since the FOV is large, several calibrator sources (say $Q$) will be
visible. The model given by \eqref{eq:scenario1} for scenario 1 can be
extended to include unknown source dependent complex gains,
\[
\bR = (\bG_1 \bK \bG_2) \bSigma_s (\bG_1 \bK \bG_2)^H \;+\; \bSigma_n\,,
\]
where $\bG_1 = \diag(\bg_1)$ represents the antenna gain, and $\bG_2 =
\diag(\bg_2)$ the source dependent complex gains, which describe the antenna
beam shape and the propagation conditions. In this model, we can merge the
unknown $\bG_2$ with $\bSigma_s$ to obtain a single unknown diagonal source
matrix $\bSigma = \bG_2 \bSigma_s \bG_2^H$, i.e.,
\[
    \bR = \bG \bK \bSigma  \bK^H \bG^H \; + \; \bSigma_n
\]
Given $\bR$, the objective is to estimate $\bG$, $\bSigma$ and $\bSigma_n$.
Here, $\bK$ is known as we know the source locations.  If there is significant
refraction, each viewing direction may pass through a different phase wedge,
causing direction dependent motion of sources (but no further deformation). In
that case, we will also have to include a parametric model for $\bK$ and solve
for the source directions (DOA estimation). This scenario is treated in, e.g.,
\cite{Weiss1995-1, Wijnholds2009-1}.

The most straightforward algorithms to solve for the unknowns are based on
alternating least squares. Assuming that a reasonably accurate starting point
is available, we can solve $\bG$, $\bSigma$ and $\bSigma_n$ in turn, keeping
the other parameters fixed at their previous estimates \cite{Wijnholds2009-1}:

\begin{itemize}
\item {\em Solve for instrument gains:}
\[
    \begin{array}{lcl}
      \hat{\mathbf{g}} & = & \underset{\mathbf{g}}{\mathrm{argmin}} 
      \| \hat{\mathbf{R}} - {\mathbf{G}} (\mathbf{K} \bSigma 
      \mathbf{K}^H) {\mathbf{G}}^H - \bSigma_n \|^2 \\
      & = & \underset{\mathbf{g}}{\mathrm{argmin}} 
      \| \vect ( \hat{\mathbf{R}} - \bSigma_n ) +\\
      & & - \diag ( \vect ( \mathbf{R}_0 ) )
      ( {\overline{\mathbf{g}} \otimes \mathbf{g} } ) \|^2
    \end{array}
\]
where $\bR_0 = \bK \bSigma \bK^H$.  This problem cannot be solved in closed
form. Alternatively, we can first solve an unstructured problem: define
$\mathbf{v} = \overline{\bg}\otimes\bg$ and solve
\[
    \hat{\mathbf{v}} = \diag ( \vect ( \mathbf{R}_0 ) 
    )^{-1} \vect ( \hat{\mathbf{R}} - \bSigma_n )
\]
or equivalently
\[
    \widehat{\mathbf{g}\mathbf{g}^H} = ( \hat{\mathbf{R}} - \bSigma_n
    ) \oslash \mathbf{R}_0.
\]
where $\oslash$ denotes a pointwise division. After this, we can do a rank-1
approximation to find $\bg$. The pointwise division can lead to noise
enhancement; this is remediated by only using the result as initial estimate
for, e.g., Gauss-Newton iteration \cite{Fuhrmann1994-1} or by formulating a
{\em weighted} least squares problem instead \cite{Wijnholds2006-1,
  Wijnholds2009-1}.

\item {\em Solve for source powers $\bsigma = \diag(\bSigma)$:}
\[\begin{array}{lcl}
  \hat\bsigma & = & \underset{\bsigma}{\mathrm{argmin}} \|
  \hat{\mathbf{R}} - \mathbf{G} \mathbf{K} {\bSigma} \mathbf{K}^H 
  \mathbf{G}^H - \bSigma_n \|_F^2 \\
  & = & \underset{\bsigma}{\mathrm{argmin}} \| \vect ( 
  ( \bRh - \bSigma_n ) - ( \mathbf{G}
  \mathbf{K} ) {\bSigma} ( \mathbf{G} \mathbf{K} )^H
  ) \|^2 \\
  & = & \underset{\bsigma}{\mathrm{argmin}} \| \vect (
  \bRh - \bSigma_n ) - \overline{( \mathbf{G}
    \mathbf{K} )} \circ ( \mathbf{G} \mathbf{K} ) {\bsigma}
  \|^2 \nonumber\\
  & = & ( \overline{\mathbf{G} \mathbf{K}} \circ \mathbf{G} \mathbf{K}
  )^\dagger \vect ( \bRh - \bSigma_n )
\end{array}
\]
\item {\em Solve for noise powers $\bsigma_n = \diag ( \bSigma_n )$:}
\begin{eqnarray*}
    \hat\bsigma_n &=& \underset{\bsigma_n}{\mathrm{argmin}} \|
	\bRh - \mathbf{G} \mathbf{K} \bSigma \mathbf{K}^H \mathbf{G}^H -
	{\bSigma_n} \|^2
    \\
    &=& 
    \diag ( \bRh - \mathbf{G} \mathbf{K}
    \bSigma \mathbf{K}^H \mathbf{G}^H ).
\end{eqnarray*}
\end{itemize}

A more optimal solution can be found by covariance matching estimation, which
provides an asymptotically unbiased and statistically efficient solution
\cite{Ottersten1998-1}. However, there is no guarantee that this will hold for
a weighted alternating least squares approach. Fortunately, the simulations in
\cite{Wijnholds2009-1} suggest that it does for this particular problem, even
if the method is augmented with weighted subspace fitting \cite{Viberg1991-1,
  Viberg1991-2}.

The first step of this algorithm is closely related to the self-calibration
(SelfCal) algorithm \cite{cornwell81,pearson84} widely used in the radio
astronomy literature, in particular for solving Scenario 1 and 2. In this
algorithm, $\bR_0$ is a reference model, obtained from the best known map at
that point in the iteration.

An alternative implementation is Field-Based Calibration
\cite{cotton04spie}. Assuming the instrumental gains have been corrected for,
an image based on a short time interval is made. The apparent position shifts
of the strongest sources are related to ionospheric phase gradients in the
direction of each source. These ``samples'' of the ionosphere are interpolated
to obtain a phase screen model over the entire field of view.  This can be
regarded as image plane calibration. The method is limited to the regime where
the ionospheric phase can be described by a linear gradient over the array.

For the MWA currently a real time calibration method based on "peeling" is
being investigated \cite{Mitchell2008-1}. In this method of successive
estimation and subtraction calibration parameters are obtained for the
brightest source in the field. The source is then removed from the data, and
the process is repeated for the next brightest source.  This leads to a
collection of samples of the ionosphere, to which a model phase screen can be
fitted.

\subsection*{Scenario 4}

This scenario is the most general case and should be applied to large arrays
with a wide field of view such as LOFAR and SKA. In this case, each station
beam sees a multitude of sources, each distorted by different ionospheric
gains and phases. The data model for the resulting direction-dependent
calibration problem is
\[
    \bR = \bA \bSigma_s \bA^H + \bSigma_n \; = \; ({\bG} \odot \bK) 
    \bSigma_s ({\bG} \odot \bK)^H+ \bSigma_n   \,,
\]
where $\bG = [\vg_1, \cdots, \vg_Q]$ is now a full matrix; $\bSigma_s$ and
$\bK$ are known. $\bG$ and $\bSigma_n$ are unknown.  Without making further
assumptions, the solution is ambiguous: the gains are not identifiable.

This problem is discussed in \cite{art_noordam_2002_1} and studied in more
detail in \cite{Tol2007-1}.  Possible assumptions that may lead to
identifiability are:
\begin{itemize}
\item {\em Bootstrapping from a compact core.} The planned geometry of LOFAR
  and SKA includes a central core of closely packed stations.  Under suitable
  conditions, these can be calibrated as under Scenario 3, giving a starting
  point for the calibration of the other stations.
\item {\em Exploiting the different time and frequency scales.} Suppose we
  have a number of covariance observations $\bR_{k,m}$, for different
  frequencies $f_k$ and time intervals $m$. The matrix $\bK = \bK_{k,m}$ is
  varying over frequency and time, whereas the instrumental gains are
  relatively constant. This can be exploited to suppress contaminating sources
  by averaging over $\bK_{k,m}$ while correcting the delays towards the
  calibrator sources.
\item {\em Modeling the gain matrix $\bG = \bG_{k,m}$.}  The gain matrix can
  be approximated by a low order polynomial model in $k$ and $m$, leading to a
  reduction in the number of unknowns. As basis functions for the polynomials
  we can use the standard basis, or Zernike polynomials (often used in
  optics), or a Karhunen-Loeve basis derived from the predicted covariance
  matrix (see Box ``Ionospheric calibration'').
\item {\em Successive estimation and subtraction or "peeling".}  In this
  method a distinction is made between the instrumental gains and ionospheric
  gains based on considerations such as temporal stability and frequency
  dependence. Sources are estimated and removed from the data in an iterative
  manner. This leads to a collection of samples to which a global model of the
  ionosphere or station beam can be fitted.
\end{itemize}
A complete calibration method that incorporates many of the above techniques
was recently proposed in \cite{Intema:2009}, and was successfully tested on a
number of 74 MHz fields observed by the VLA. The behavior of this method at
lower frequencies and / or on baselines longer than a few tens of kilometers
still needs investigation.

\section{Compound element calibration}
\label{sec:compoundcal}

Compound elements are used in very large aperture arrays to keep the number of
correlator inputs manageable, and as focal plane arrays to increase the FOV of
dishes. In either case, each compound element produces a superposition of
antenna signals, $\bx(n)$, at its output port $y(n)$ during normal operation,
i.e.,
\begin{equation*}
y(n) = \vw^H \vx(n)
\end{equation*}
where $\vw$ are the beamformer weights. Compound elements therefore require a
separate calibration measurement before or after the observation, as only
$y(n)$ is available and no antenna specific information can be derived from
this superposition. Compound elements should thus be designed to be stable
over the time scales of a typical observation.

Initial system characterization is often done in an anechoic chamber. In these
measurements, the response $y(n)$ of the compound element to a test probe is
recorded, while varying the beamformer $\bw$ \cite{Hampson1999-1}.  The
measurements $y_1(n), y_2(n), \cdots, y_N(n)$ are stacked in a vector $\vy$
while the corresponding weights $\vw_1, \vw_2, \cdots, \vw_N$ are stacked in a
matrix $\mW$. The complete series can be summarized as
\begin{equation*}
\vy(n) = \mW^H \bG \vk s(n) 
\end{equation*}
where $s(n)$ is the known input signal, $\vk$ is the phase vector describing
the geometrical delays due to the array and source geometry and $\bG = \diag
\left ( \bg \right )$ contains the instrumental gains. The gains of the
individual elements stacked in $\bg$ are then easily derived. An attractive
feature of this method is that it can also be used in the field using a
stationary reference antenna \cite{Wijnholds2003-1}.

Calibration of an aperture array of compound elements is discussed in the
previous section.  Regarding its use as a focal plane array in a dish, there
are a number of differences, mostly because the dish projects an image of the
sky within its FOV onto the FPA, whereas an aperture array measures the
complex field distribution over the aperture itself.

The goal of FPA beamforming is to maximize the signal-to-noise ratio in an
observation \cite{Warnick2006-1}.  This involves a trade-off between
maximizing the gain in the direction of the source and minimization of the
total noise in the system.  A very intuitive approach to maximize the gain
towards the source is conjugate field matching. For this calibration method,
the array response $\va_p$ is measured for a strong point source for each of
the $P$ compound beams that will be formed by the FPA. The weights of the
$p$-th compound beam are chosen such that $\vw_p = \va_p$. Since the dish
forms an image of the point source on the FPA, most of the energy is
concentrated on a few elements. Conjugate field matching thus assigns very
high weights to a few elements and the noise of these elements will therefore
dominate the noise in the observation. If one of these elements has a poor
noise performance, conjugate field matching does not lead to the maximum
signal-to-noise ratio in the observation. The measurement on a strong point
source should therefore be augmented with a measurement on an empty sky to
obtain the noise covariance matrix of the array.  This step allows proper
weighting of the receiver paths to optimize the signal-to-noise ratio of the
actual observation.  An excellent overview of FPA signal processing is
provided in \cite{Jeffs2008-1}.

\section{Future challenges}
\label{sec:outlook}

Instruments like LOFAR and SKA will have an unprecedented sensitivity that is
two orders of magnitude higher (in the final image) than current instruments
can provide.  The increased sensitivity and large spatial extent requires new
calibration regimes, i.e., scenarios 3 and 4, which are dominated by direction
dependent effects.  Research in this area is ongoing.  Although many ideas are
being generated, only a limited number of new calibration approaches have
actually been tested on real data. This is hardly surprising since only now
the first of these new instruments are producing data.  Processing real data
will remain challenging and drive the research in this area of signal
processing.  Apart from the challenges discussed already throughout the text,
some of the remaining challenges are as follows.
\begin{itemize}
\item {\em The sky.}  Because of the long baselines that are part of the new
  instruments, many sources which appear point-like to existing instruments,
  will be resolved. This means that they cannot be treated as point sources,
  but should be modeled as extended sources using, e.g., shapelets
  \cite{Refreiger2003-1}.  The new instruments will have wide frequency bands,
  so that the source structure may change over the observing band.  For these
  reasons the source models will have to be more complicated than currently
  assumed.  At the same time, due to the increased sensitivity, many more
  sources will be detected and will have to be processed.  This will not only
  affect the calibration of the instruments, but also the imaging and
  deconvolution.

  Because of their high sensitivity the new instruments are capable of
  detecting very weak sources, but they will have to do so in the presence of
  all the strong sources already known. Some of those strong sources may not
  even be in the FOV, but may enter through the primary beam side lobes.

\item {\em The instrument.} Both aperture arrays and dishes with FPAs have
  primary beams that are less stable than single pixel primary beams from
  dishes. The primary beam is time dependent (if it is not fixed on the sky)
  and varies with frequency and over the different stations or FPA
  systems. These beam pattern variations have a negative impact on the
  achievable image quality. Calibrating for these beams is a real challenge,
  as is the correction for such beams during imaging.

\item {\em The atmosphere.} For low frequency instruments, ionospheric
  calibration is a significant challenge.  Current algorithms have been
  shown to work for baselines up to a few tens of kilometers and
  frequencies as low as 74~MHz.  However, for baselines of a few
  hundreds up to a few thousands of kilometers and frequencies down to,
  say, 10~MHz these algorithms may not be valid.

\item {\em Polarization purity.} Calibration and imaging have to take the
  full polarization of the signal into account. The primary beam of an
  instrument introduces instrumental polarization due to the reception
  properties of the feeds. If the feeds do not track the rotation of the sky,
  as is the case in any radio telescope that does not have an equatorial
  mount, the instrumental polarization varies over the observation. The
  ionosphere alters the polarization of the incoming electromagnetic waves as
  well due to Faraday rotation. These effects require calibration and
  correction with high accuracy.

\item {\em The large number of elements.}  Classical radio telescopes have at
  most a few tens of receivers (WSRT has 14 dishes and the VLA has 27).
  Instruments like LOFAR, MWA and EMBRACE will have about $10^4$ receiving
  elements while SKA is envisaged to have over $10^6$ signal paths.  The
  corresponding increase in data volumes will require sophisticated
  distributed signal processing schemes and algorithms that can run on
  suitable high performance computing hardware.

\item {\em Equations and unknowns.} It is clear that in order to deal with
  these challenges more complicated models are needed, which in turn contain
  more unknowns that need to be extracted from the data. The increase in the
  number of stations will yield more equations, but this may not be
  enough. Modeling of the time-frequency dependence of parameters by a
  suitable set of basis functions will decrease the amount of unknowns that
  need fitting.

\item {\em Interference mitigation.} The radio frequency spectrum is rather
  crowded, and it is expected that many observations will be contaminated by
  (weak or strong) RFI. Array signal processing techniques can be used to
  suppress interference, e.g., by active null steering or covariance matrix
  filtering \cite{laj99astro,aj04sam}.  For LOFAR and SKA, no techniques have
  been proposed yet. Research from cognitive radio and compressive sensing may
  be very relevant for interference avoidance.
\end{itemize}

Finding suitable answers to these challenges will be of critical importance
for the next generation of instruments.

\section*{Acknowledgments}

This work was supported in part by NWO-STW (The Netherlands) under grant
number DTC.5893. The Netherlands Institute for Research in Astronomy (ASTRON)
is an institute of the Netherlands Organization for Scientific Research (NWO).

Credits for photos and artists impressions: Karl Jansky's antenna, Fig.\
\ref{fig:instruments}$(a)$: National Radio Astronomy Observatory (NRAO,
www.nrao.edu); Effelsberg, Fig.\ \ref{fig:instruments}$(b)$: Stefan Wijnholds;
WSRT, Fig.\ \ref{fig:instruments}$(c)$: Netherlands Institute for Radio
Astronomy (ASTRON, www.astron.nl); VLA, Fig.\ \ref{fig:instruments}$(d)$: NRAO
(www.vla.nrao.edu); ALMA, Fig.\ \ref{fig:instruments}$(e)$: ALMA observatory
(www.almaobservatory.org); concept layout of LOFAR, Fig.\ \ref{fig:hierarchy}
$(a)$: ASTRON; LOFAR low-band antennas, Fig.\ \ref{fig:hierarchy}$(b)$: Menno
Norden; MWA tile, Fig.\ \ref{fig:hierarchy}$(c)$: MWA project
(www.mwatelescope.org); artist impressions SKA, Figs.\
\ref{fig:hierarchy}$(e)$ and \ref{fig:hierarchy}$(f)$: SKA project
(www.skatelescope.org); SKA demonstrator, Fig.\ \ref{fig:hierarchy}$(g)$:
ASTRON.

\section*{Authors} {\em Stefan J.  Wijnholds} (wijnholds@astron.nl) was born
in The Netherlands in 1978. He received the M.Sc.\ degree in Astronomy and the
M.Eng.\ degree in Applied Physics (both cum laude) from the University of
Groningen in 2003.  After his graduation he joined the R\&D department of
ASTRON, the Netherlands Institute for Radio Astronomy, in Dwingeloo, The
Netherlands, where he works with the system design and integration group on
the development of the next generation of radio telescopes.  Since 2006 he is
also with TU Delft, The Netherlands, where he is pursuing a Ph.D.\ degree.
His research interests lie in the area of array signal processing,
specifically calibration and imaging.

{\em Sebastiaan van der Tol} (vdtol@strw.leidenuniv.nl) was born in The
Netherlands in 1977.  He received the M.Sc.\ degree in Electrical Engineering
from TU Delft, The Netherlands, in 2004.  Since January 2004 he has been a
research assistant with the same institute, where he is pursuing a Ph.D.\
degree in Electrical Engineering.  His current research interests include
array signal processing and interference mitigation techniques for large
phased array radio telescopes.

{\em Ronald Nijboer} (rnijboer@astron.nl) was born in The Netherlands in 1971.
He received the Ph.D.\ degree in 1998 from the Vrije Universiteit, Amsterdam.
From 1998 till 2004 he worked in the field of aeroacoustics with the National
Aerospace Laboratory (NLR).  In 2004 he started working for ASTRON, where at
present he is leading the Computing group.  His research interests include
calibration and imaging algorithms.

{\em Alle-Jan van der Veen} (a.j.vanderveen@tudelft.nl) was born in The
Netherlands in 1966.  He received the Ph.D.\ degree (cum laude) from TU Delft
in 1993.  Throughout 1994, he was a postdoctoral scholar at Stanford
University.  At present, he is a Full Professor in Digital Signal Processing
at TU Delft.  He is the recipient of a 1994 and a 1997 IEEE Signal Processing
Society (SPS) Young Author paper award, and was an Associate Editor for IEEE
Tr.\ Signal Processing (1998--2001), chairman of IEEE SPS Signal Processing
for Communications Technical Committee (2002-2004), Editor-in-Chief of IEEE
Signal Processing Letters (2002-2005), Editor-in-Chief of IEEE Transactions on
Signal Processing (2006-2008), and member-at-large of the Board of Governors
of IEEE SPS (2006-2008).  He is currently member of the IEEE SPS Awards Board
and IEEE SPS Fellow Reference Committee.  He is a Fellow of the IEEE.

\bibliography{refs}

\end{document}